\documentclass[preprint,pre,amsmath,amssymb,showpacs]{revtex4}

\usepackage{epsfig}
\usepackage{epstopdf}
\usepackage{graphics}
\usepackage{multirow}
\usepackage{bm}
\begin{document}
\title{Continuous surface force based lattice Boltzmann equation method for simulating thermocapillary flow}
\author{Lin Zheng$^{1}$}
\email[Corresponding author:\quad]{lz@njust.edu.cn}
\author{Song Zheng$^2$}
\author{Qinglan Zhai$^3$}
\affiliation{1School of Energy and Power Engineering, Nanjing University of Science and Technology, Nanjing 210094, P.R. China}
\affiliation{2School of Mathematics and Statistics, Zhejiang University of Finance and Economics, Hangzhou Zhejiang 310018, P.R. China}
\affiliation{3 School of Economics Management and Law, Chaohu University, Chaohu 238000, P.R. China}
\begin{abstract}

In this paper, we extend a lattice Boltzmann equation (LBE) with continuous surface fore (CSF) to simulate thermocapillary flows. The model is designed on our previous CSF LBE for athermal two phase flow, in which the interfacial tension forces and the Marangoni stresses as the results of the interface interactions between different phases are described by a conception of CSF. In this model, the sharp interfaces between different phases are separated by a narrow transition layers, and the kinetics and morphology evolution of phase separation would be characterized by an order parameter visa Cahn-Hilliard equation which is solved in the frame work of LBE. The scalar convection-diffusion equation for temperature field is also solved by thermal LBE. The models are validated by thermal two layered Poiseuille flow, and a two superimposed planar fluids at negligibly small Reynolds and Marangoni numbers for the thermocapillary driven convection, which have analytical solutions for the velocity and temperature. Then thermocapillary migration of two dimensional deformable droplet are simulated. Numerical results show that the predictions of present LBE agreed with the analytical solution/other numerical results.

\end{abstract}
\maketitle

\section{Introduction}

With the rapid development of the materia and aviation technology, the transport mechanism of interfacial thermodynamics under microgravity/zero gravity or in microfluidic system is one of the hot topics in space science. In the microgravity environment or the microfluidic devices, the effect of gravity is greatly eliminated or even disappeared, then different transports of the interface dynamics are emerged. When the system has a nonuniform temperature distribution,  there has a temperature gradient along the interface, which caused to a variation of the surface tension along the interface (the surface tension generally decreases with the increased temperature for most fluids).  This variable surface tension force could lead to a viscous stress, which could induce the fluid's motion from a hotter region to a colder region. This phenomena is known as thermocapillary (Marangoni) convection, which plays a dominant role in microgravity \cite{Subra} or microfluidic devices \cite{Darhuber}.

As one of the interesting investigations of the thermocaplliary convection, the migration of an unconfined spherical droplet/bubble has been investigated extensively \cite{Young,Subra}. In Ref. \cite{Young}, Young et al first derived an analytical formulation for the terminal velocity of unconfined non deformable drop with a linear temperature profile in the creeping flow limit. Since then, there have been numerous subsequent experimental or numerical
studies to investigate such phenomenon \cite{Subra}.
As we known, experimental investigations of the thermocapillary migration of droplet/bubble are hampered by gravitational effects which tend to mask the thermocapillary effect on terrestrial. To reduce such effects, the drop tower, sounding rockets, and aboard space shuttles are the basic ways to get a short time microgravity environment for investigating thermocapillary convection,  while a long time microgravity experiment in space station is an expensive and crucial way, which depended on the aviation program in the whole world \cite{Kawamura}. Although experimental investigations could help to understand the phenomena of thermocapillary flows in microgravity/microfluidic devices, it is still difficult to precisely measure the local temperature and flow fields during the transport process of a droplet/bubble.

On the other hand, numerical method has been viewed as a scientific method for the fluid dynamics, which has been successfully applied to the thermocapillary flows \cite{Sim,Hu,Du,Zhenga,Hariri,Yin,Liu,Liu1}. However, an efficient and precise description of phase interaction or its interfacial dynamics model is still a challenging task. In literature, there are generally two clarifications of numerical methods for simulating thermocapillary flow: one is single phase Navier-Stokes equations (NSE) based numerical method \cite{Sim,Hu,Du,Zhenga}; another is the two phase NSE based numerical method \cite{Hariri,Yin,Liu,Liu1}. In the former method, the physical problem is mainly focused on dynamics of one phase, and the thermocapillary effect is entreatment through the interface boundary conditions, which is usually used to some simplified thermocapillary convection problems.  In the latter one, the detailed physical phenomena of the interface dynamics in both phases could be observed, and there is no need to implement the interface boundary conditions throughout the computation except for the sharp interface method. It is well understood that the interface dynamics of the two phase flow is just the result of molecular interactions between different phases. Thus, if we could design a model that could correctly description of such interaction process at microscopic level, the corresponding interface dynamics could be obtained at macroscopic level. The lattice Boltzmann method (LBE) is just one of the mesoscopic methods, which could be applied to model such interaction process \cite{He,Lee,Liu,Liu1,Zheng}.


In this paper, we will extend previous continuous surface force (CSF) LBE to the thermocapillary flow, and the effect of the Marangoni force is included through the CSF formulation. The evolution of interface is governed by the Cahn-Hilliard equation (CHE), which is solved by LBE, and a thermal LBE is derived from the kinetic theory for solving the scalar convection-diffusion energy equation. The rest of this paper is organized as follows. In Sec II, a continuous surface force formulation of LBE model is presented, and a LBE model for temperature field is proposed in Sec. III, then some numerical simulations are conducted to validate the models in Sec. IV, and finally a brief conclusion is given in Sec. V.

\section{LBE with continuous surface force }

In general, surface tension is a function of local temperature in thermal multiphase system, so the effect of tangential gradient of the surface tension should be included in the CSF formulation, and the governing equation for the momentum could be written as
\begin{equation}
\partial_t(\rho\bm{u})+\nabla\cdot(\rho\bm{uu})=-\nabla p+\nabla\cdot\bm{S}+\bm{F},
\label{Eq1}
\end{equation}
where $\rho$ is the fluid density, $\bm{u}$ is the velocity, $p$ is the hydrodynamic pressure, $\bm{S}$ is the viscous stress term and the interface force $\bm{F}$ in Eq. (\ref{Eq1}) is given as \cite{Landau}
\begin{equation}
\bm{F}=-\sigma\kappa\delta\bm{n}+\nabla_s\sigma\delta
\label{Eq2}
\end{equation}
where $\sigma$ is the surface tension, $\kappa$ is the total curvature, $\delta$ is a regularized delta function, $\bm{n}$ is the ourward pointing unit normal vector, and $\nabla_s=(\bm I-\bm{nn})\cdot\nabla$ is the surface gradient operator. The first term on the right hand side of Eq. (\ref{Eq2}) is the normal surface tension force, and the second is the tangential (Marangoni) force which is the result of the nonuniform surface tension. Alternately, one may write $\bm F$ in a stress formulation
\begin{equation}
\bm{F}=\nabla\cdot[(\bm{I-nn})\sigma\delta]
\label{aEq2}
\end{equation}

With this interface force formulation in Eq. (\ref{Eq2}), a diffuse interface formulation of $\bm{F}$ could be written as
\begin{equation}
\bm{F}=(-\sigma\nabla\cdot\bm{n}\bm{n}+\nabla_s\sigma)\epsilon\alpha|\nabla c|^2
\label{Eq3}
\end{equation}
where $\bm{n}=\nabla c/|\nabla c|$ with $c$ the order parameter, $\epsilon$ is a small parameter related to the interface thickness and $\alpha$ is a normalized constant to be determined later. Comparing with Eqs. (\ref{Eq2}) and (\ref{Eq3}), the curvature $\kappa$ relate to the unit normal vector $\bm{n}$ as $\kappa=\nabla\cdot\bm{n}$ and the regularized delta function $\delta$ relate to the order parameter as $\delta=\epsilon\alpha|\nabla c|^2$.

In thermocapillary flow, the flow is driven by surface tension force which is a function of the temperature. For simplicity, we assume that the relation of the surface tension to the temperature is a linear relation in present work
\begin{equation}
\sigma=\sigma_0+\sigma_T(T-T_0),
\label{aEq3}
\end{equation}
where $\sigma_0$ is the surface tension at the reference temperature $T_0$, $\sigma_T = \partial\sigma/\partial T$ is the rate of change of interfacial tension with temperature, and $T$ is local temperature.

With the formulations of Eqs (\ref{Eq3}) and (\ref{aEq3}), we can derive a similar incompressible LBE model with CSF for the fluid flow \cite{He,Lee,Zheng}

\begin{equation}
\begin{array}{rr}
f_i(\bm x+\bm\xi_i\delta t, t+\delta t)-f_i(\bm x, t)=-\omega(f_i-f^{(eq)}_i)
+\delta t(1-\frac{\omega}{2})[\bm{F}\cdot(\bm\xi_i-\bm u)\Gamma_i(\bm u)\\
+(\bm\xi_i-\bm u)\cdot \nabla(\rho c^2_s)(\Gamma_i(\bm u)-\Gamma_i(0))],
\label{Eq12}
\end{array}
\end{equation}
where $f_i$ is the density distribution function, $\bm\xi_i$ is the molecular velocity, $\delta t$ is the time step, $\omega=1/\tau_f$ is the relaxation rate with $\tau_f$ the relaxation time, and $f^{(eq)}_i$ is equilibrium density distribution function which given as
\begin{equation}
f^{(eq)}_i=\omega_i\left[p+\rho c^2_s\left(\frac{\bm\xi_i\cdot\bm u}{c^2_s}+\frac{1}{2}\left((\frac{\bm\xi_i\cdot\bm u}{c^2_s})^2-\frac{\bm u^2}{c^2_s}\right)\right)\right],
\label{Eq5}
\end{equation}
and $\Gamma_i(\bm u)$ is given as
\begin{equation}
\Gamma_i(\bm u)=\omega_i\left\{1+\frac{\bm\xi_i\cdot\bm u}{c^2_s}+\frac{1}{2}\left[\left(\frac{\bm\xi_i\cdot\bm u}{c^2_s}\right)^2-\frac{\bm u^2}{c^2_s}\right]\right\},
\label{Eq6}
\end{equation}
where $\omega_i$ is the weight coefficient depending on the number of discrete velocity $\bm\xi_i$, $c_s$ is sound speed. The dynamic pressure and velocity defined by the velocity moments of the density distribution function are given by

\begin{equation}
p=\sum_i f_i+\frac{\delta t}{2}\bm{u}\cdot\nabla\rho c^2_s, ~~~~\rho c^2_s\bm u=\sum_i\bm\xi_if_i+\frac{\delta t}{2}\bm{F}
\label{Eq13}
\end{equation}

Through the Chapmann-Enskog (CE) multiscale analysis for Eq. (\ref{Eq12}), the following governing equations could be obtained
\begin{eqnarray}
\nabla\cdot \bm u &=&0,\\
\partial_t(\rho\bm{u})+\nabla\cdot(\rho\bm{uu})&=&-\nabla p+\nabla\cdot\bm{S}+\bm{F}.
\end{eqnarray}
where the viscous stress $\bm S=\eta(\nabla\bm u+ \bm u\nabla)$ with viscosity $\eta=\rho c^2_s(\tau_f-1/2)\delta t$.

In the phase field theory, the kinetics and morphology evolution of phase separation is characterized by CHE via an order parameter $c$. It is usually used to identify the two phase region, where $c=c_1$ occupied by fluid 1, and $c=c_2$ occupied by fluid 2. The mixing free energy of this fluid for the isothermal system without the solid boundaries can be written as
$$E=\int [E_0+\frac{\epsilon^2}{2}|\nabla c|^2]d\Omega,$$
where $E_0$ is a bulk energy, which is related to the bulk chemical potential by $\mu_0=\partial_c E_0$. In the phase field theory, $E_0$ can be approximated by
$E_0(c)=\beta(c-c_1)^2(c-c_2)^2$ with $\beta$ a constant coefficient, $c_1$ and $c_2$ are respectively the corresponding order parameter at fluid 1 and fluid 2. For planar interface at $z=0$ in an equilibrium system, the distribution of the order parameter has the following analytical solution
$$c(z)=\frac{c_1+c_2}{2}+\frac{c_1-c_2}{2}tanh(\frac{z}{2\sqrt{2}\epsilon}),$$
where $z$ is distance to the interface, to match the surface tension of the sharp interface model, the constant $\alpha$ in Eq. (\ref{Eq3}) should choose as
$$\epsilon\alpha\int^\infty_{-\infty} |c(z)|^2dz=1.$$
Therefore, with above equation and set $c_1=0$ and $c_2=1$, we could get $\alpha=6\sqrt{2}$ and the phase interface at $c=0.5$.

In this paper, we apply recently proposed LBE for the CHE, and the evolution equation of the order parameter is \cite{Zheng}

\begin{equation}
g_i(\bm x+\bm\xi_i\delta t, t+\delta t)-g_i(\bm x, t)=-\omega_h(g_i-g^{(eq)}_i)
+\delta t(1-\frac{\omega_h}{2})\left[\frac{c(\bm{\xi}_i-\bm u)}{\rho c^2_s}\cdot(\bm F-\nabla p)\right]\Gamma_i(\bm u),
\label{Eq27}
\end{equation}
where $\omega_h=1/\tau_g$ with $\tau_g$ the relaxation time, and $g^{(eq)}_i$ is equilibrium distribution function of order parameter, which is given as \cite{Zheng}
\begin{equation}
g^{(eq)}_i=\omega_i\left\{H_i+c\left[\frac{\bm\xi_i\cdot\bm u}{c^2_s}+\frac{1}{2}\left((\frac{\bm\xi_i\cdot\bm u}{c^2_s})^2-\frac{\bm u^2}{c^2_s}\right)\right]\right\},
\label{Eq21}
\end{equation}
with the coefficient $H_i$ given as

\begin{equation}
H_i=\left\{\begin{array}{ll} [c-(1-\omega_0)\Gamma\mu/c^2_s]/\omega_0, &\textrm{\emph{i}=0}\\
\Gamma\mu/c^2_s, &\textrm{\emph{i}}>0\\
\end{array}\right.
\label{Eq22}
\end{equation}
and the order parameter could be calculated by
\begin{equation}
c=\sum_i g_i.
\label{Eq28}
\end{equation}
Through CE expansion, the time evolution of the order parameter could be derived \cite{Zheng}
\begin{equation}
\partial_t c +\nabla\cdot(\bm uc)=\nabla\cdot(M\nabla\mu),
\label{Eq16}
\end{equation}
where $M=\delta t\Gamma(\tau_g-1/2)$ is the mobility, and $\mu=\mu_0-\epsilon^2\nabla^2 c$ is the chemical potential.

\section{LBE model for temperature field }

In this section, we will derive a two phase thermal LBE for temperature field, the model start from the kinetic theory of Boltzmann equation with BGK collision operator \cite{BGK}

\begin{equation}
\partial_t \bar{f}_i+\bm\xi_i\cdot\nabla \bar{f}_i=-\frac{\bar{f}_i-\bar{f}^{(eq)}_i}{\tau}+\frac{\bm{\bar{F}}\cdot(\bm\xi_i-\bm u)}{\rho c^2_s}\bar{f}^{(eq)}_i,
\label{aEq4}
\end{equation}
where $\bar{f}_i$ is the density distribution function, $\tau$ is the relaxation time, $\bm{\bar{F}}=\nabla\rho c^2_s-\nabla p+\bm{F}$ and $\bar{f}^{(eq)}_i$ is equilibrium density distribution function which given as
\begin{equation}
\bar{f}^{(eq)}_i=\frac{\rho}{(2\pi RT)^{D/2}}\text{exp}\left[\frac{(\bm{\xi}_i-\bm u)^2}{2RT}\right],
\label{aEq5}
\end{equation}

Similar with previous work \cite{He1}, we introduce a new distribution function as
\begin{equation}
h_i=\frac{(\bm{\xi_i-u})^2}{2}\bar{f}_i,
\label{aEq6}
\end{equation}
then the following evolution for $h_i$ could be derived
\begin{equation}
\partial_t h_i+\bm\xi_i\cdot\nabla h_i=-\frac{h_i-h^{(eq)}_i}{\tau_h}+\bar{f}_iq_i+\frac{\bm{\bar{F}}\cdot(\bm\xi_i-\bm u)}{\rho c^2_s}h^{(eq)}_i,
\label{aEq7}
\end{equation}
where $q_i=(\bm{\xi}_i-\bm u)\cdot(\partial_t \bm u+\bm{\xi}_i\cdot \nabla\bm u)$ is related to the pressure work and viscous dissipation term, if this two terms are negligible in energy equation, $\bar{f}_iq_i$ in Eq. (\ref{aEq7}) could be dropped, then Eq. (\ref{aEq7}) could be simplified as
\begin{equation}
\partial_t h_i+\bm\xi_i\cdot\nabla h_i=-\frac{h_i-h^{(eq)}_i}{\tau_h}+\frac{\bm{\bar{F}}\cdot(\bm\xi_i-\bm u)}{\rho c^2_s}h^{(eq)}_i,
\label{aEq8}
\end{equation}
and through the Hermit expansion and neglected high order of $O(\bm u^3)$, $h^{(eq)}_i$ could be written as \cite{Shi}
\begin{equation}
h^{(eq)}_i=\rho c_vT\Gamma_i(\bm u),
\label{aEq9}
\end{equation}
and the temperature is calculated as
 \begin{equation}
\rho c_vT=\sum_i h_i,
\label{aEq10}
\end{equation}
where $c_v$ is the specific ratio of heat with constant volume.

Integrating Eq. (\ref{aEq8}) along the characteristic line from time $t$ to $t + \delta t$ and using trapezoidal discretization rule, we
introduce the following new distribution function
\begin{equation}
\bar{h}_i=h_i-\frac{\delta t}{2}\left[-\frac{h_i-h^{(eq)}_i}{\tau_h}+\frac{\bm{\bar{F}}\cdot(\bm\xi_i-\bm u)}{\rho c^2_s}h^{(eq)}_i\right],
\label{aEq11}
\end{equation}
and then we could obtain the evolution equation for the temperature field as

\begin{equation}
\bar{h}_i(\bm x+\bm\xi_i\delta t, t+\delta t)-\bar{h}_i(\bm x, t)=-\omega_h(\bar{h}_i-h^{(eq)}_i)
+\delta t(1-\frac{\omega_h}{2})\frac{\bm{\bar{F}}\cdot(\bm\xi_i-\bm u)}{\rho c^2_s}h^{(eq)}_i,
\label{aEq12}
\end{equation}
where $\omega_h=2\delta t/(2\tau_h+\delta t)$. The temperature defined by the velocity moments of the new distribution function is given by

 \begin{equation}
\rho c_vT=\sum_i \bar{h}_i,
\label{aEq13}
\end{equation}

Through the CE expansion for Eq. (\ref{aEq12}) (See Appendix A for details), the following governing equation could be obtained
\begin{equation}
\partial_t(\rho c_vT)+\nabla\cdot(\rho c_v \bm u T)=\nabla\cdot\lambda\nabla T,
\label{aEq14}
\end{equation}
where the conductivity $\lambda=\rho c^2_s\tau_h\delta t$.

\section{Numerical simulations}

In this section, we will conduct three benchmark problems to validate the proposed LBE. The test problems include a thermal layered Poiseuille flow driven by external force, and two superimposed planar fluids at negligibly small Reynolds and Marangoni numbers for the thermocapillary driven convection, which have analytical solutions for the velocity and temperature profiles, and a two dimensional (2D) deformable droplet migration by the temperature gradient.

\subsection{Thermal layered Poiseuille flow}

\begin{figure}
\includegraphics[width=0.5\textwidth,height=0.25\textwidth]{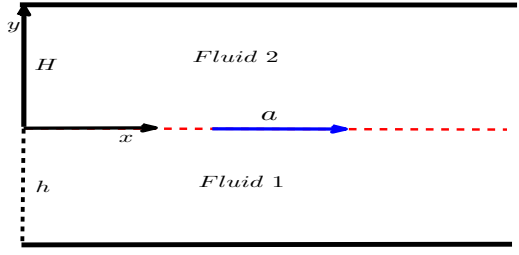}
\caption{Sketch of layered Poiseuille flow in a 2D channel}\label{Fig1}
\end{figure}

\begin{figure}
\includegraphics[width=2.5in,height=2.0in]{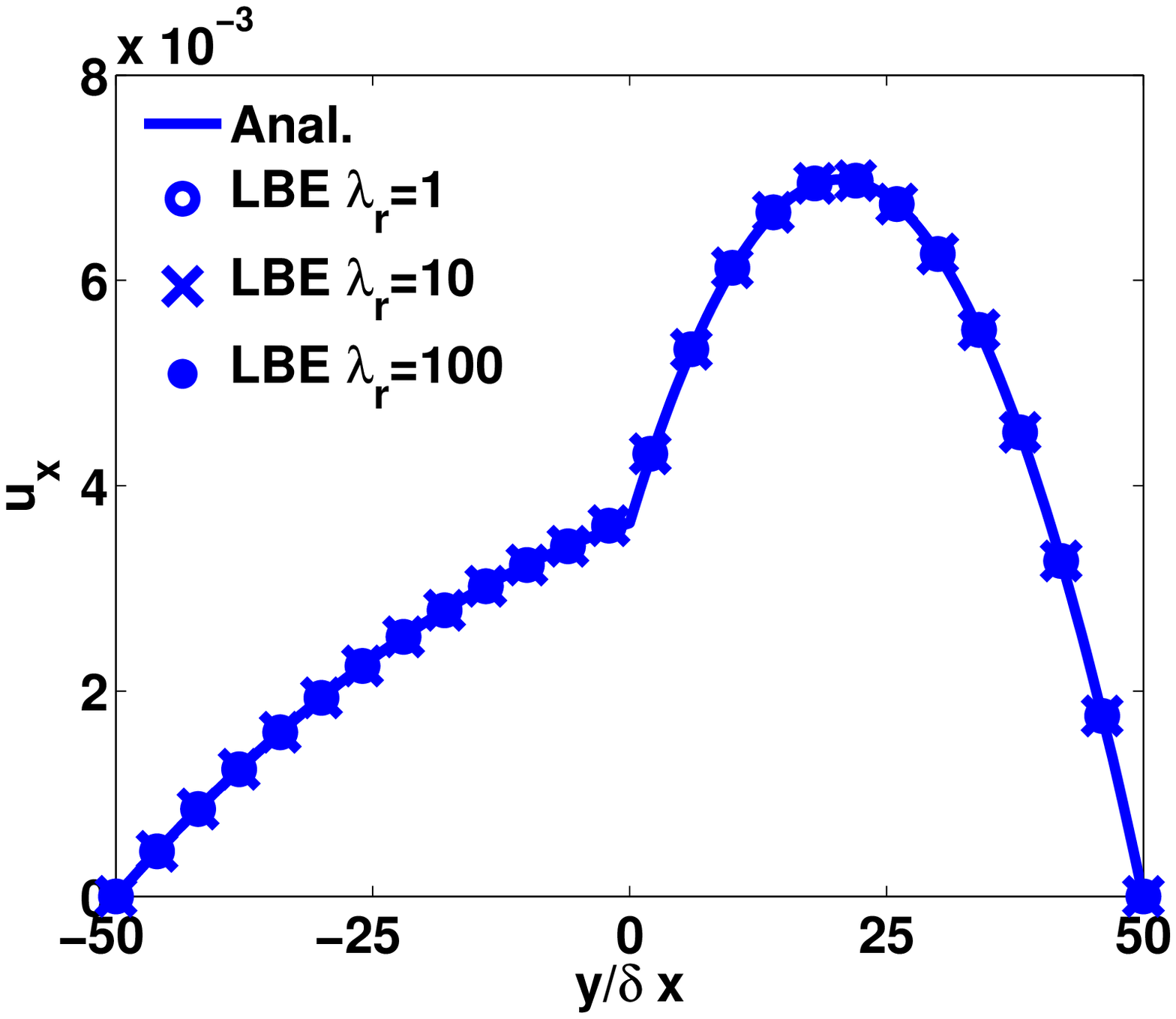}
\includegraphics[width=2.5in,height=2.0in]{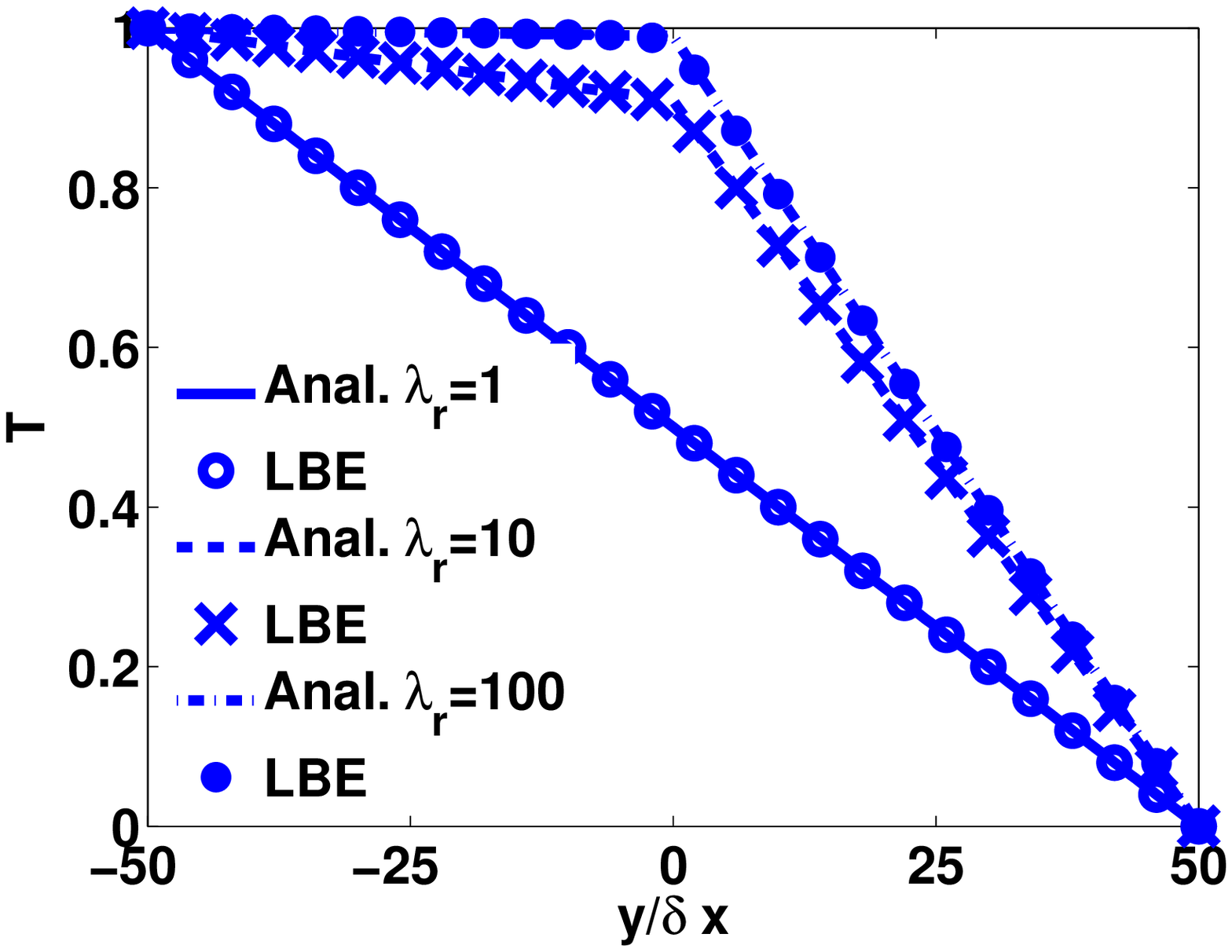}
\caption{Velocity (the left figure) and temperature (the right figure) profiles of layered Poiseuille flow with $\eta_r=10$, $\lambda=1, 10, 100$, $M=0$. Solid and dashed lines are analytic solutions, circles, crosses and dots are the numerical results}\label{Fig3}
\end{figure}

The first test problem is thermal layered Poiseuille flow between two infinite
plates driven by a constant external force $\bm a=(a_x, 0)$, which provides a good benchmark for validating the present thermal LBE. The computational domain is $0\leq x\leq L$ and $-h\leq y\leq H$ with two layers. For simplicity, the fluid in the domain has the same density $\rho$, and the fluid with viscosity $\eta_1$ is at lower layer $(-h\leq y<0)$, while the upper layer $(0<y\leq H)$ with a viscosity $\eta_2$, the physical configuration is shown in Fig. \ref{Fig1}. The nonslip boundary condition is applied to the solid boundaries, a higher temperature $T_h$ is fixed at the top wall while a lower temperature $T_c$ is applied to the bottom wall, periodic boundary condition is applied to the $x$ direction.  The fluid velocity, temperature, stresses and heat fluxes across the interface should be continuous, as the flow is sufficiently slow that there is no deformation occur at the interface, and the surface tension was assumed to be constant, then analytical solutions of steady $x$ component of velocity profile $u_x$ with constant surface tension for this problem is
\begin{figure}
\includegraphics[width=0.3\textwidth,height=0.25\textwidth]{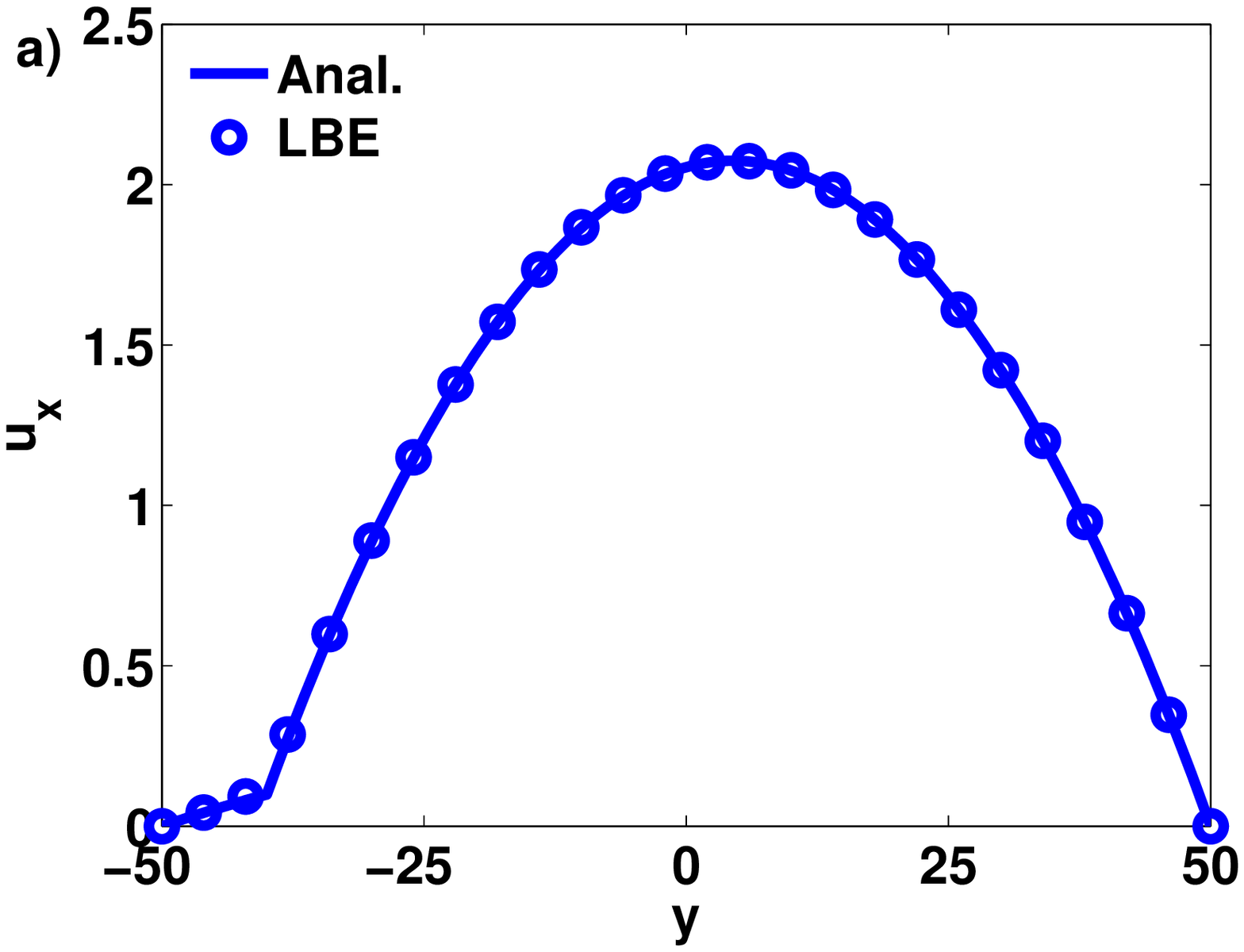}
\includegraphics[width=0.3\textwidth,height=0.25\textwidth]{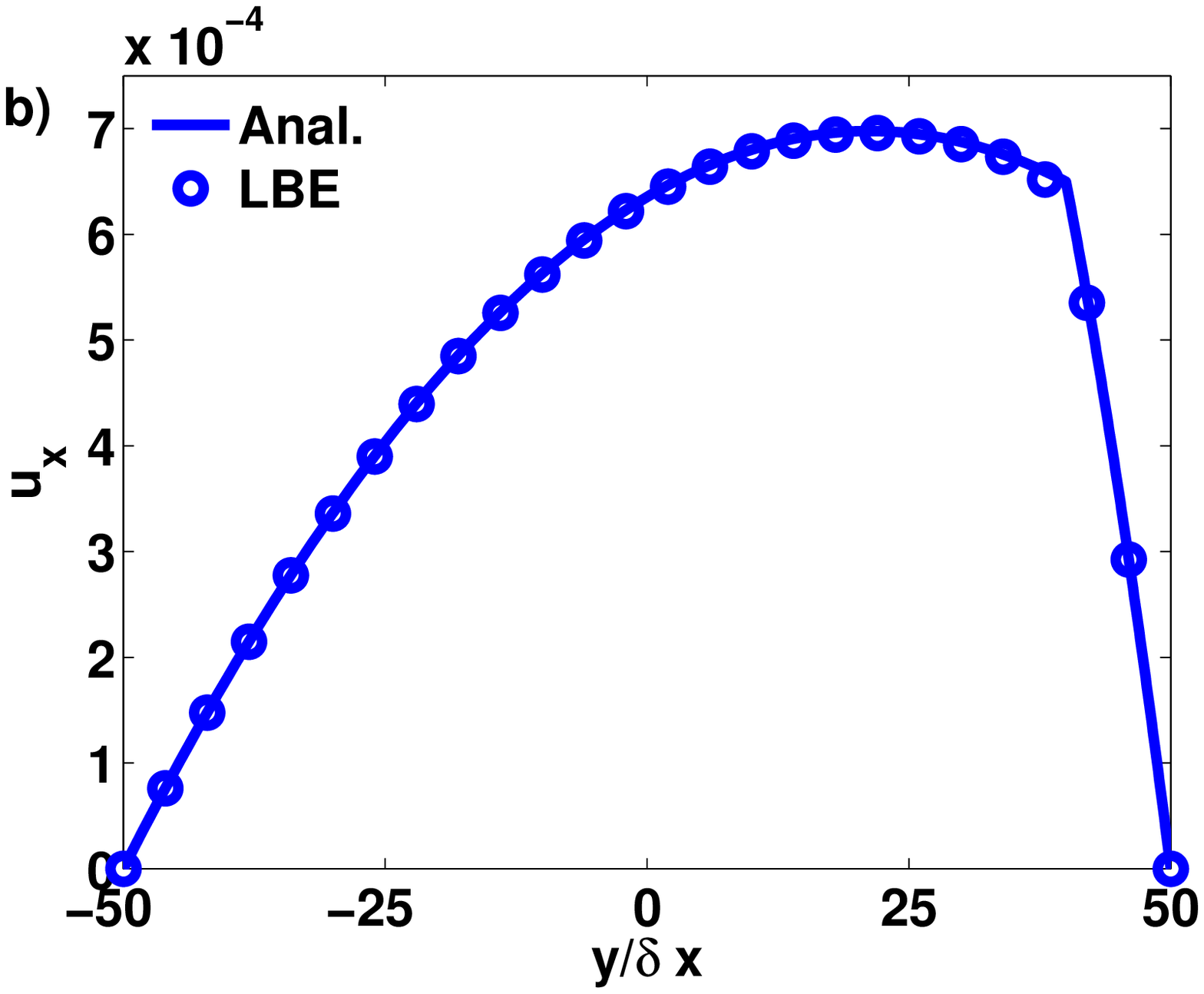}
\includegraphics[width=0.3\textwidth,height=0.25\textwidth]{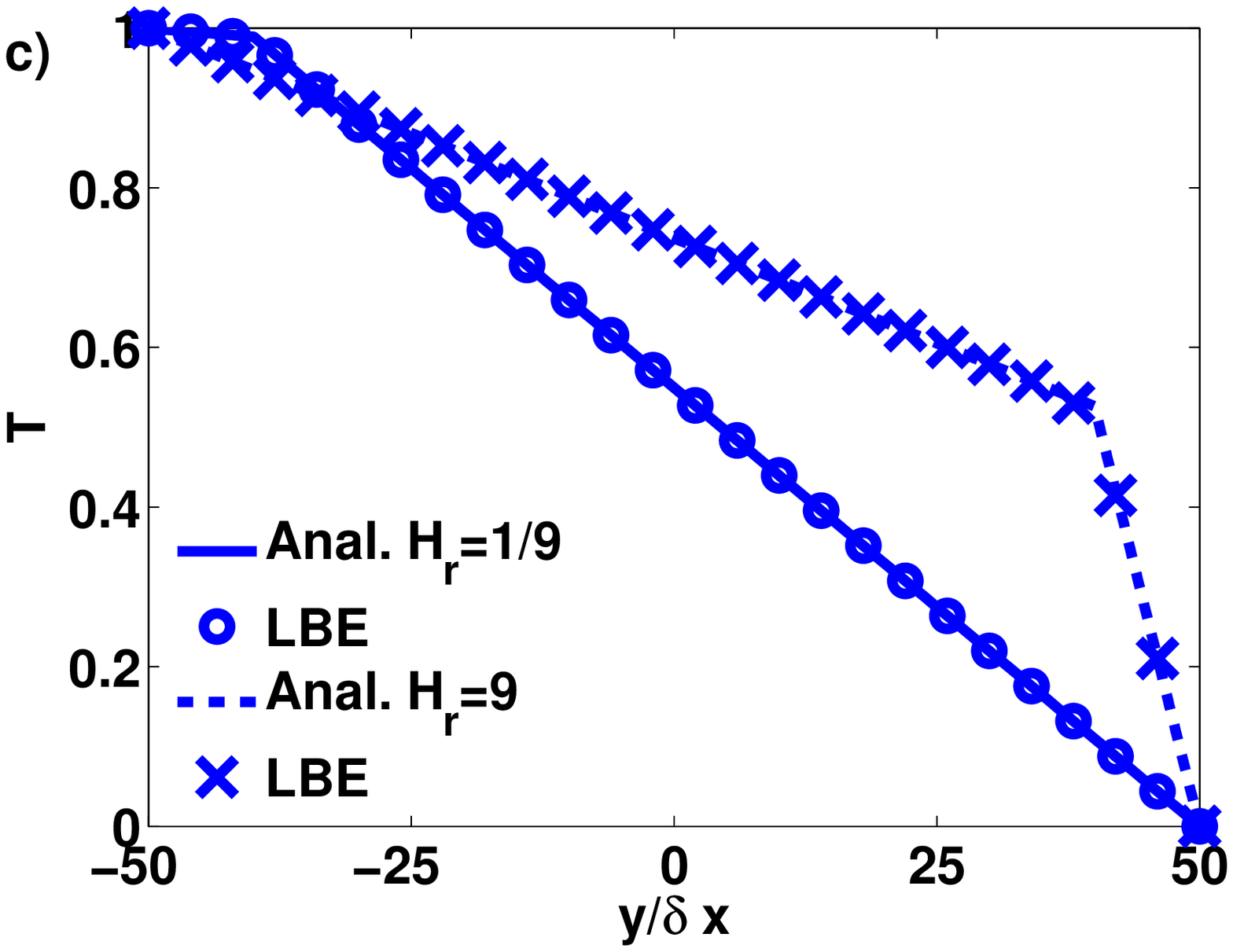}
 \caption{Velocity (a) $H_r=1/9$, b) $H_r=9$) and temperature (c)) profile of layered Poiseuille flow with $\eta_r=10$, $M=0$. Solid and dashed
lines are analytic solutions, circles and crosses are the numerical results}\label{Fig4}
\end{figure}

\begin{equation}
u_x=\left\{\begin{array}{ll} \frac{\rho a_xh^2}{2\eta_1}\left[-(\frac{y}{h})^2+\left(\frac{H^2/h^2\eta_1 -\eta_2}{H/h\eta_1+\eta_2}\right)\frac{y}{h}+\frac{(1+H/h)H\eta_1}{\eta_1H+\eta_2h}\right], &-h\leq y\leq 0\\
\frac{\rho a_xH^2}{2\eta_2}\left[-(\frac{y}{H})^2+\left(\frac{\eta_1 -h^2/H^2\eta_2}{\eta_1+h/H\eta_2}\right)\frac{y}{H}+\frac{(1+h/H)h\eta_2}{\eta_1H+\eta_2h}\right], &~~~0\leq y\leq H
\label{Eq32}
\end{array}\right.
\end{equation}
and the temperature profile $T$ for this problem could be obtained as
\begin{equation}
T=\left\{\begin{array}{ll} \frac{\lambda_2(T_h-T_c)}{\lambda_1H+\lambda_2h}y+\frac{\lambda_1HT_c+\lambda_2hT_h}{\lambda_1H+\lambda_2h}, &-h\leq y\leq 0\\
\frac{\lambda_1(T_h-T_c)}{\lambda_1H+\lambda_2h}y+\frac{\lambda_1HT_c+\lambda_2hT_h}{\lambda_1H+\lambda_2h}, &0\leq y\leq H
\label{Eq33}
\end{array}\right.
\end{equation}

From Eqs. (\ref{Eq32}) and (\ref{Eq33}), the velocity profile is related to viscosity ratio ($\eta_r=\eta_1/\eta_2$), and the layer thickness ratio ($H_r=h/H$), and the temperature is related to the conductivity ratio ($\lambda_r=\lambda_1/\lambda_2$), $H_r$. In simulations, a $40\times100$ mesh has been employed, periodic boundary condition is applied to the $x$ direction, and nonslip boundary condition is used at solid boundaries. The other parameters in LBE are chosen as $\rho=1.0, \beta=0.25, \eta_1=0.1, \lambda_1=0.1, \epsilon=1$ and Reynolds number Re=$\rho u_1h/\eta_1=1$ with $u_1=\rho a_xh^2/2\eta_1$. In Fig. \ref{Fig3}, numerical results for different values of $\lambda_r$=1, 10, 100 with $\eta_r=10$ and $H_r=1$ are compared, the figures shown that numerical results of LBE agreed well with the analytical solutions. We also consider the effect of the layer thickness ratios $H_r=1/9, 9$ in Fig. \ref{Fig4} with $\eta_r=10$, $\lambda_r=10$, the results shown that present LBE could give a satisfied result for the thermal layered flow.

\subsection{ Thermocapillary flow with two superimposed planar fluids }
\begin{figure}
\includegraphics[width=0.5\textwidth,height=0.25\textwidth]{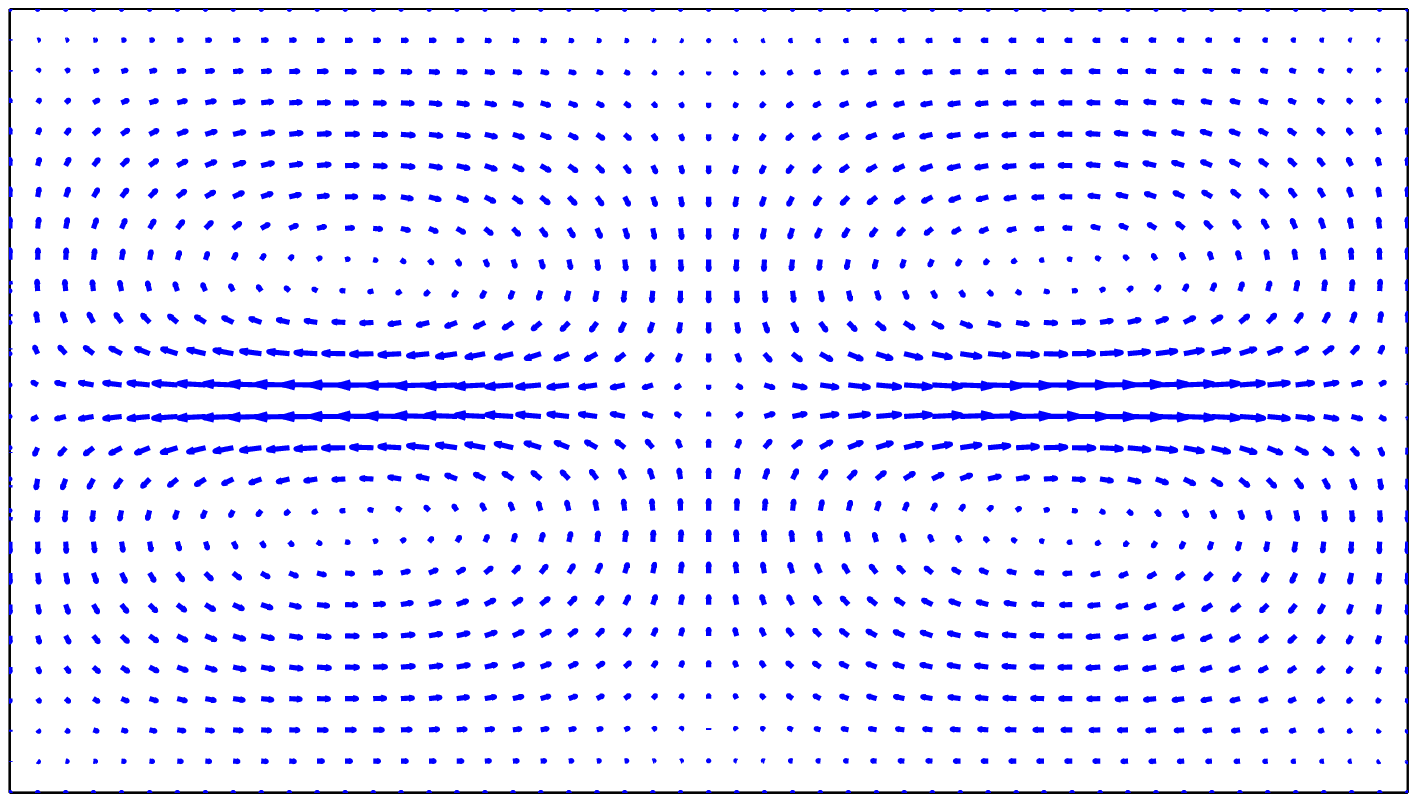}%
\includegraphics[width=0.5\textwidth,height=0.25\textwidth]{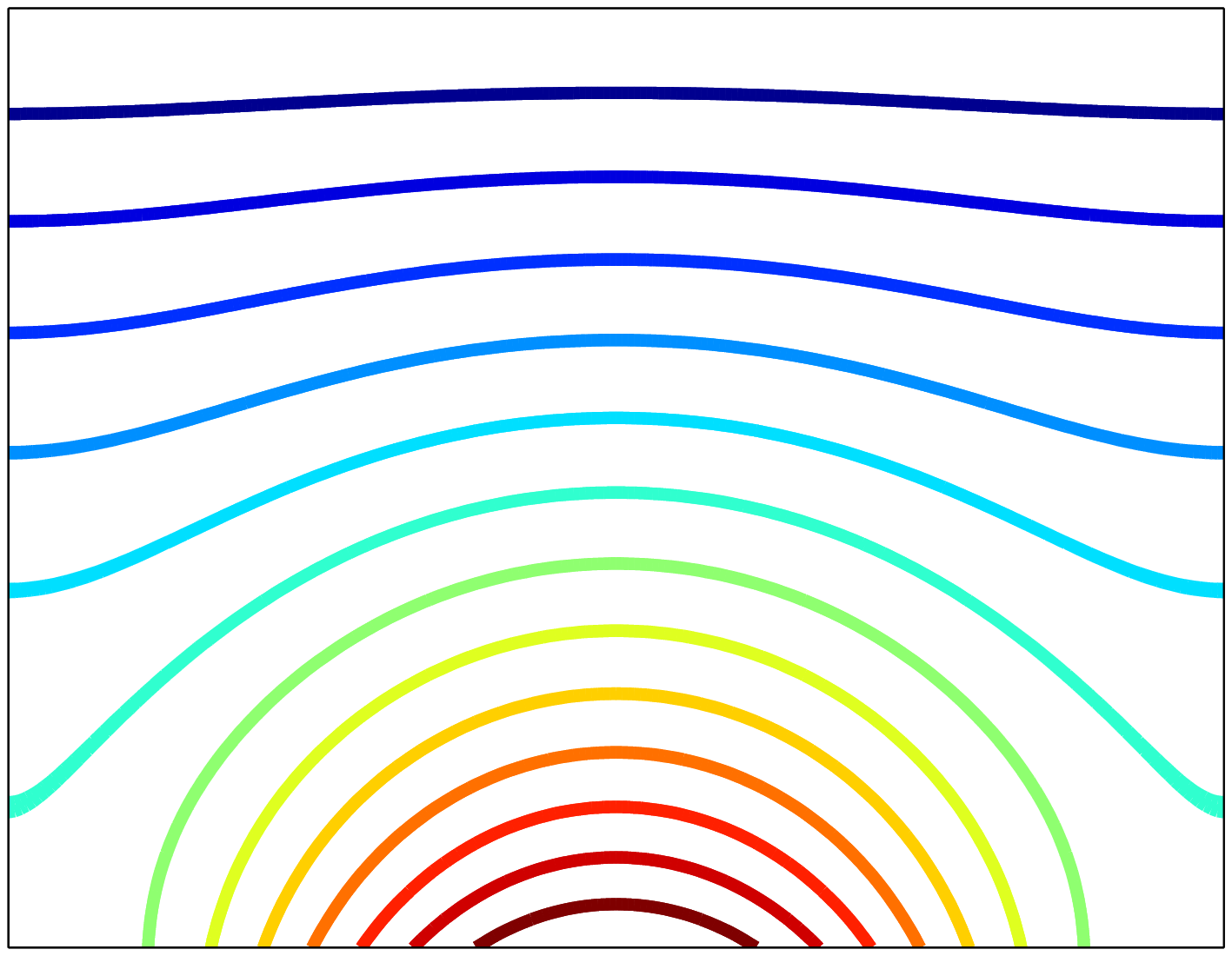}
\includegraphics[width=0.5\textwidth,height=0.25\textwidth]{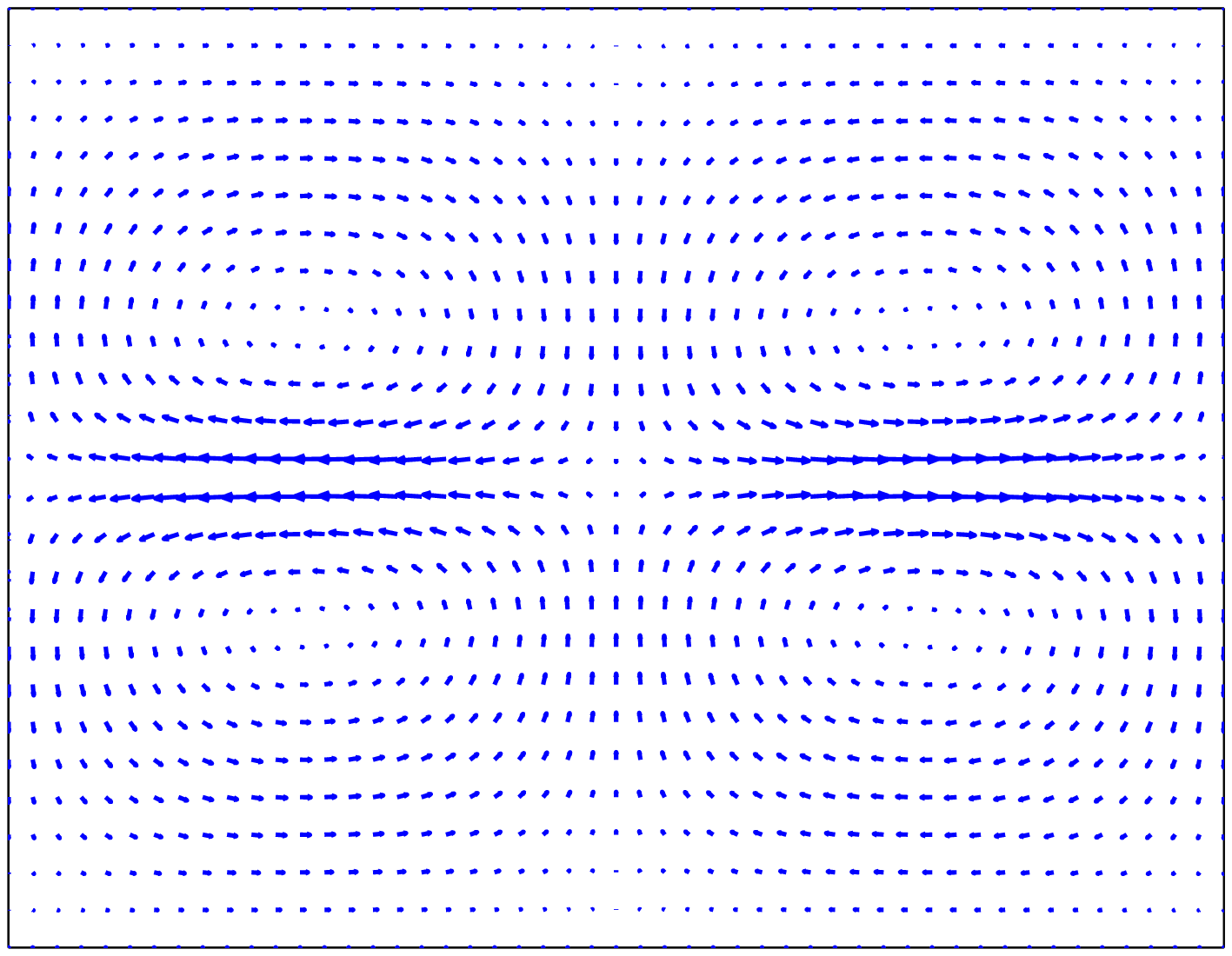}%
\includegraphics[width=0.5\textwidth,height=0.25\textwidth]{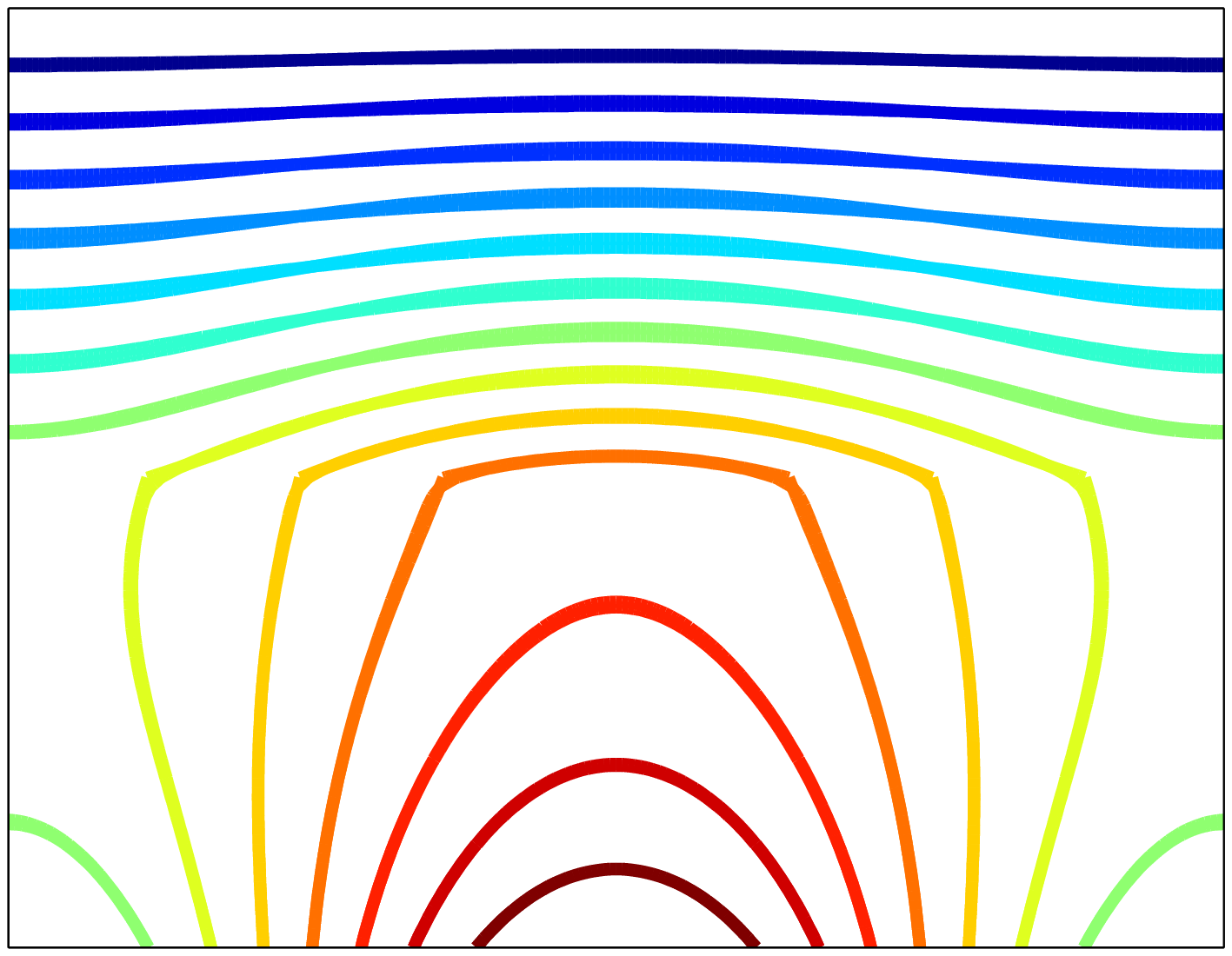}
\caption{Velocity fields and isothermal lines for $\lambda_r=1$ (top) and $\lambda_r=10$ (bottom).}\label{Fig5}
\end{figure}

Two superimposed planar fluids driven by temperature gradient provide another good benchmark for validating the present LBE. The computational domain is $-L/2\leq x\leq L/2$ and $-h\leq y\leq H$ with two layers. The fluid with density $\rho_1$, viscosity $\eta_1$ and thermal conductivity $\lambda_1$ is at upper layer $(0<y\leq H)$, while the lower layer $(-h\leq y<0)$ with a density $\rho_2$, viscosity $\eta_2$ and thermal conductivity $\lambda_2$, the physical configuration is shown in Fig. \ref{Fig1}. The nonslip boundary condition is applied to the solid boundary and periodic boundary condition is applied to the $x$ direction.  For the thermal boundary conditions, we assume that the upper wall with an uniform temperature, while a sinusoidal temperature is imposed at the lower wall as
\begin{equation}
T(H,x)=T_c,
\label{aEq15}
\end{equation}
and
\begin{equation}
T(-h,x)=T_h+T_r\text{cost}(kx),
\label{aEq16}
\end{equation}
where $0<T_r<T_c<T_h$, and $k=2\pi/L$ is a wave
number.

\begin{figure}
\includegraphics[width=0.33\textwidth,height=0.3\textwidth]{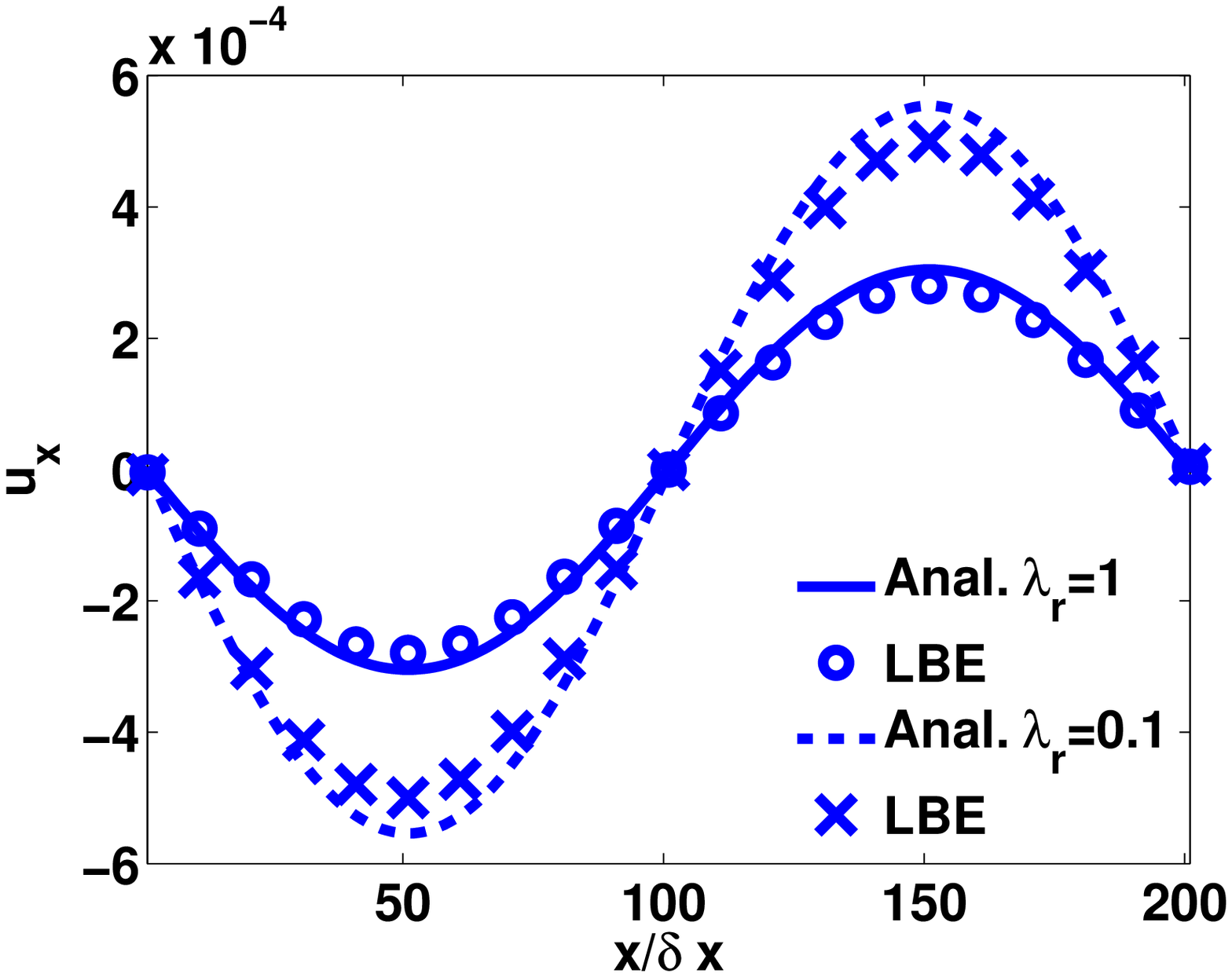}%
\includegraphics[width=0.33\textwidth,height=0.3\textwidth]{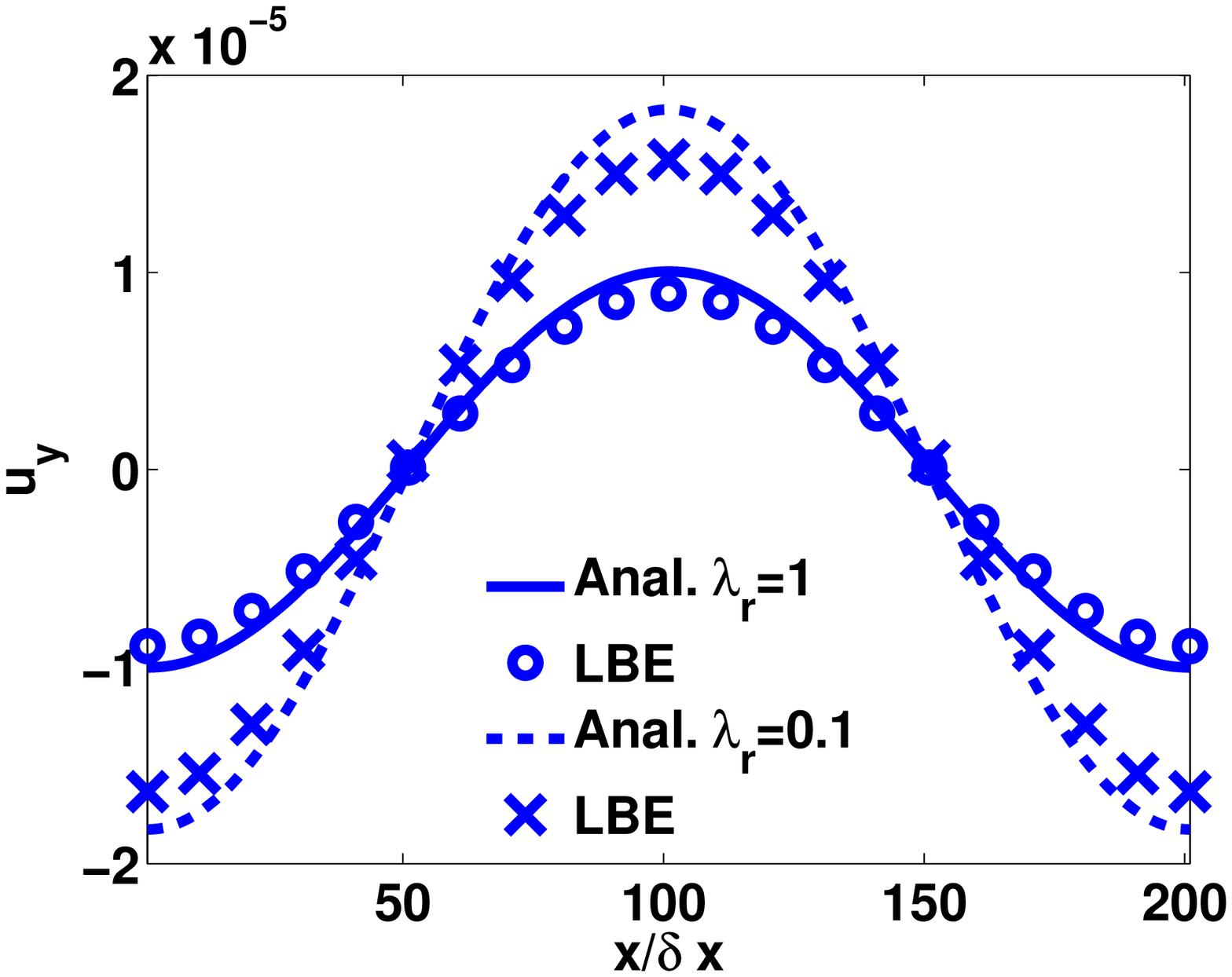}%
\includegraphics[width=0.33\textwidth,height=0.3\textwidth]{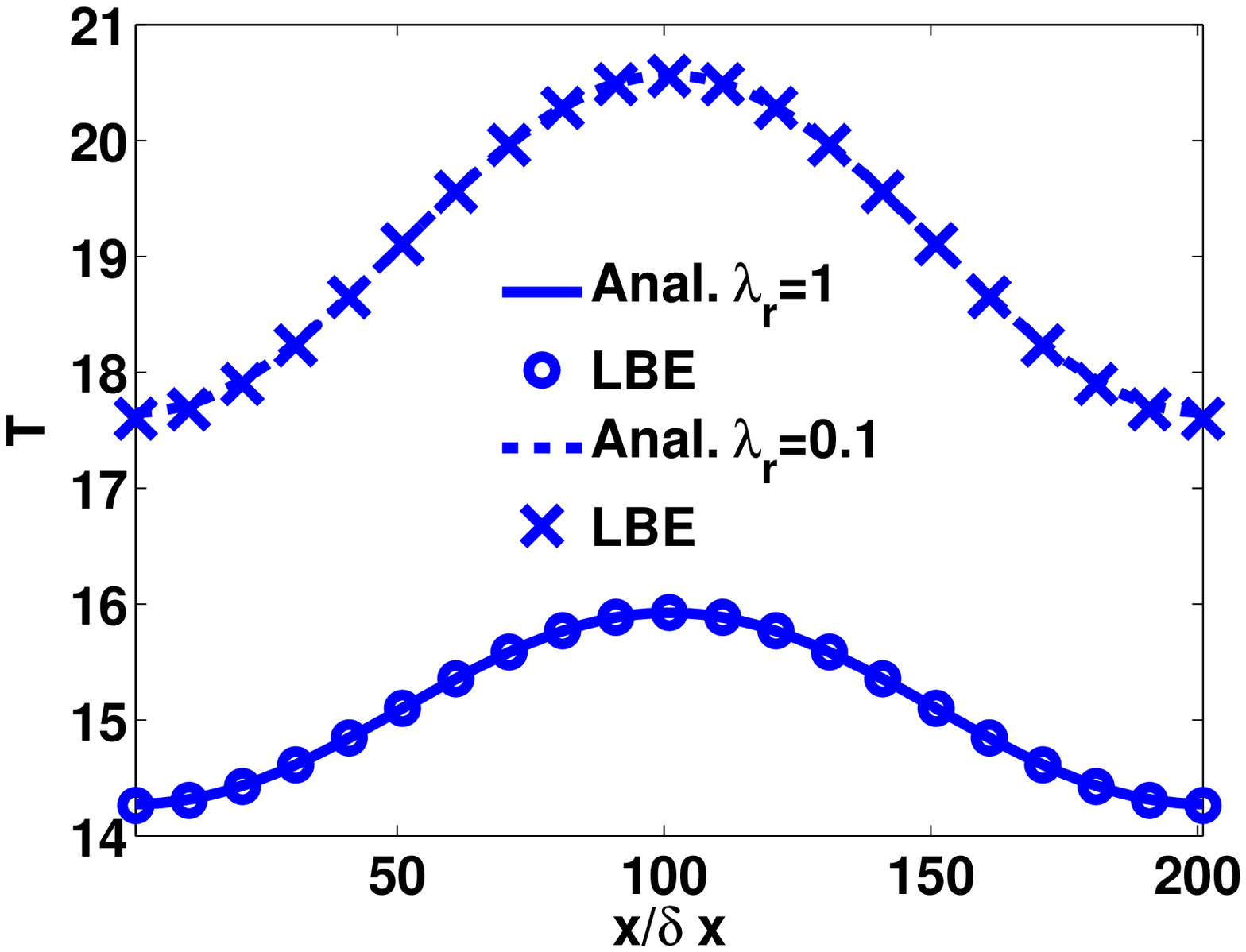}
 \caption{Velocity component and temperature profiles along the centre line of of the domain in the $x$ direction.}\label{Fig6}
\end{figure}

Assuming Re$\ll$1, Ma$\ll$1, and Ca$\ll$1, the interface is thought to remain flat, and the momentum and energy equations can be simplified to be linear, then following the procedure of Pendse and Esmaeeli, the analytical solutions for the velocity and temperature
fields were obtained as \cite{Pendse}
\begin{eqnarray}
u_x(x,y)&=&U_{max}\{[C^H_1+k(C^H_2+C^H_3y)]\text{cosh}(ky)+(C^H_3+kC^H_1y)\text{sinh}(ky)\}\text{sin}(kx),\\
u_y(x,y)&=&-kU_{max}[C^H_1y\text{cosh}(ky)+(C^H_2+C^H_3y)\text{sinh}(ky)]\text{cos}(kx),\\
T(x,y)&=&\frac{(T_c-T_h)y+\lambda_r T_ch+T_hH}{H+\lambda_r h}+T_rf(\alpha,\beta,\lambda_r)\text{sinh}(\alpha-ky)\text{cos}(kx),
\end{eqnarray}
for the upper fluid 1, and
\begin{eqnarray}
u_x(x,y)&=&U_{max}\{[C^h_1+k(C^h_2+C^h_3y)]\text{cosh}(ky)+(C^h_3+kC^h_1y)\text{sinh}(ky)\}\text{sin}(kx),\\
u_y(x,y)&=&-kU_{max}[C^h_1y\text{cosh}(ky)+(C^h_2+C^h_3y)\text{sinh}(ky)]\text{cos}(kx),\\
T(x,y)&=&\frac{\lambda_r(T_c-T_h)y+\lambda_r T_ch+T_hH}{H+\lambda_r h}+T_rf(\alpha,\beta,\lambda_r)[\text{sinh}(\alpha)\text{cosh}(ky)\nonumber\\
&&-\lambda_r\text{sinh}(ky)\text{cosh}(\alpha)]\text{cos}(kx),
\end{eqnarray}
for the lower fluid 2, the parameters in above equations could be given as
\begin{eqnarray}
\alpha=Hk;~~\beta=hk,\\
f(\alpha, \beta, \lambda_r)=[\lambda_r\text{cosh}(\alpha)\text{sinh}(\beta)+\text{sinh}(\alpha)\text{cosh}(\beta)]^{-1},\\
C^H_1=\frac{\text{sinh}^2(\alpha)}{\text{sinh}^2(\alpha)-\alpha^2};~~C^H_2=\frac{-H\alpha}{\text{sinh}^2(\alpha)-\alpha^2};~~C^H_3=\frac{2\alpha-\text{sinh}(2\alpha)}{2[\text{sinh}^2(\alpha)-\alpha^2]},\nonumber\\
C^h_2=\frac{\text{sinh}^2(\beta)}{\text{sinh}^2(\beta)-\beta^2};~~C^h_2=\frac{-h\beta}{\text{sinh}^2(\beta)-\beta^2};~~C^h_3=\frac{2\beta-\text{sinh}(2\beta)}{2[\text{sinh}^2(\beta)-\beta^2]},
\end{eqnarray}
and
\begin{equation}
U_{max}=-\frac{T_r\sigma_T}{\eta_2}g(\alpha, \beta, \lambda_r)h(\alpha, \beta, \eta_r),
\end{equation}
where
\begin{equation}
g(\alpha, \beta, \lambda_r)=\text{sinh}(\alpha)f(\alpha, \beta, \lambda_r),
\end{equation}
and
\begin{equation}
h(\alpha, \beta, \eta_r)=\frac{[\text{sinh}^2(\alpha)-\alpha^2][\text{sinh}^2(\beta)-\beta^2]}{\eta_r[\text{sinh}^2(\beta)-\beta^2][\text{sinh}(2\alpha)-2\alpha]+[\text{sinh}^2(\alpha)-\alpha^2][\text{sinh}(2\beta)-2\beta]}.
\end{equation}

\begin{figure}
\includegraphics[width=0.33\textwidth,height=0.3\textwidth]{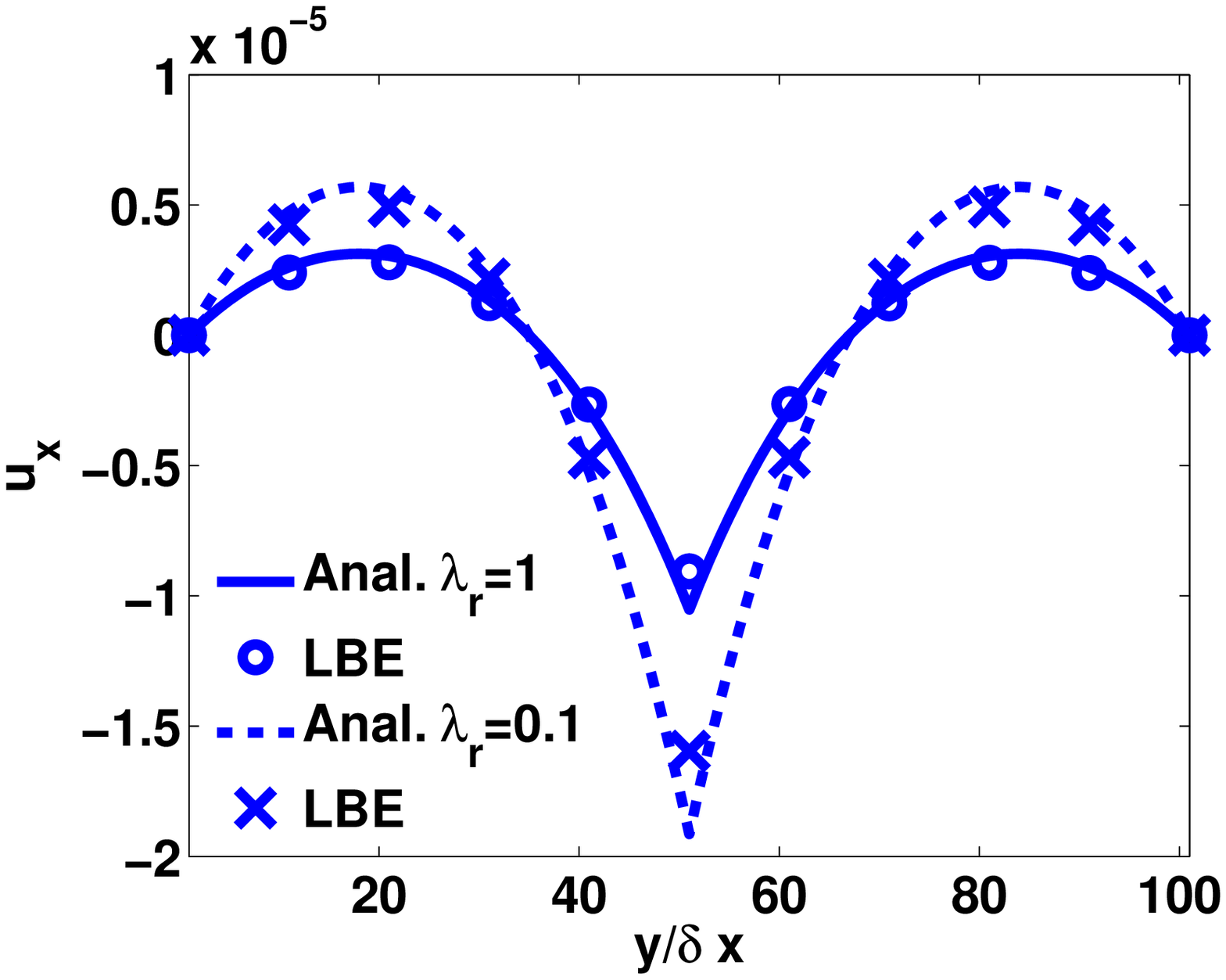}%
\includegraphics[width=0.33\textwidth,height=0.3\textwidth]{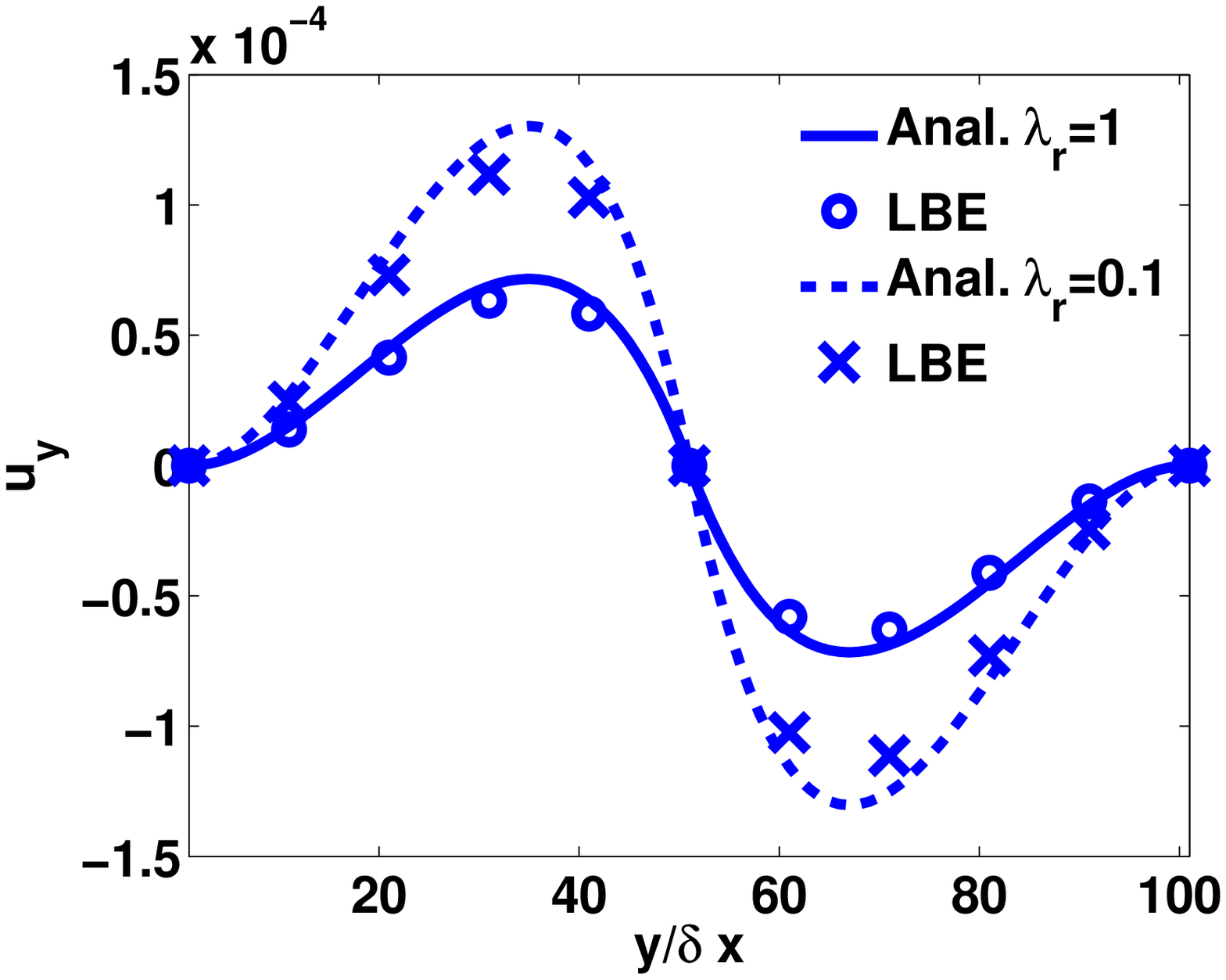}%
\includegraphics[width=0.33\textwidth,height=0.3\textwidth]{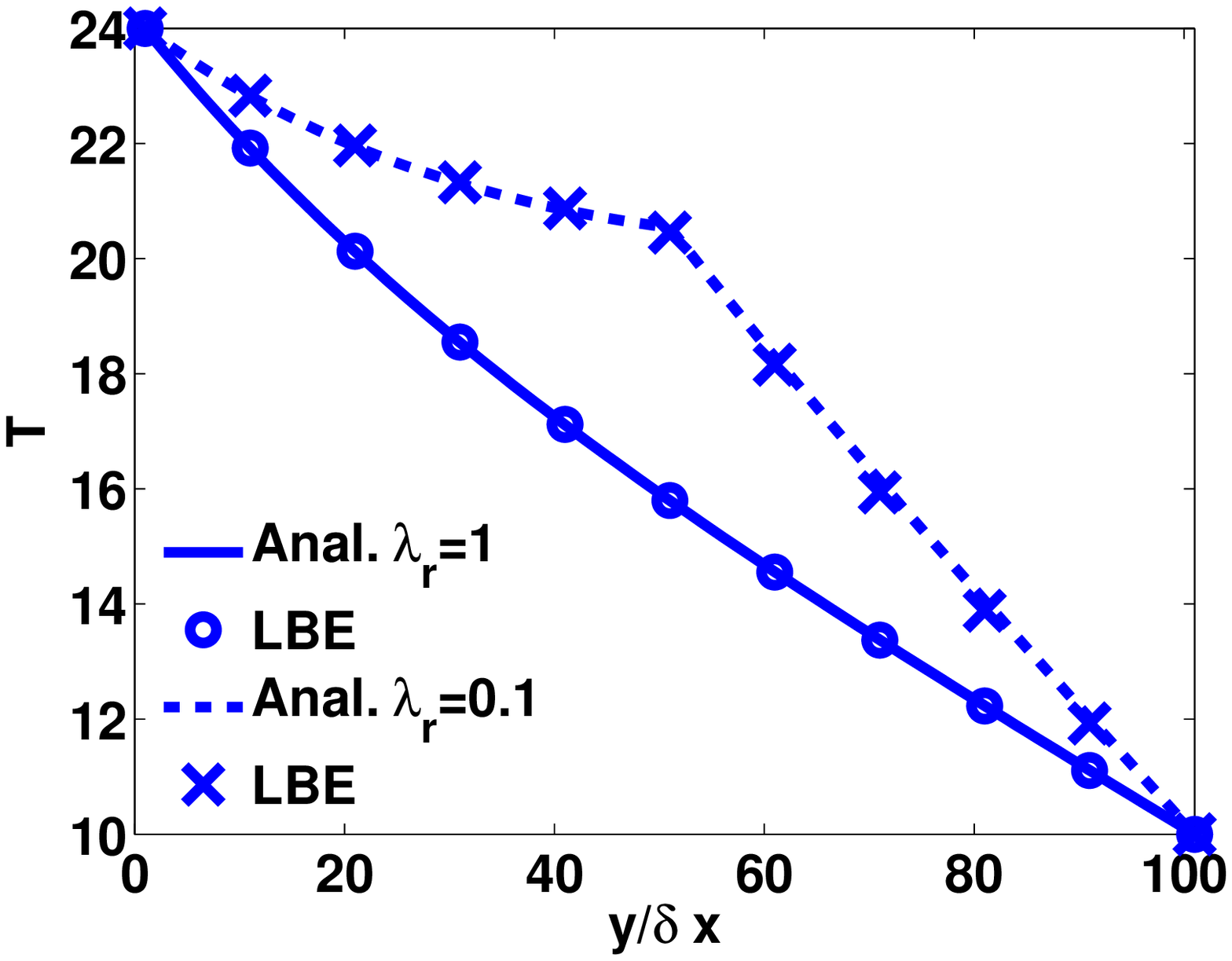}
 \caption{Velocity component and temperature profiles along the centre line of of the domain in the $y$ direction}\label{Fig7}
\end{figure}

In numerical simulation, a $200\times100$ mesh has been employed with the fluid layer $H=h$, the periodic boundary condition is applied to the $x$ direction, nonslip boundary condition is used at solid boundaries , and the value of wall temperature is implemented by Eqs. (\ref{aEq15}) and (\ref{aEq16}). The fluid properties and the parameters in LBE are chosen as $\sigma_T=-5\times10^{-4}, \sigma_0=2.5\times10^{-2}, T_0=T_c=10, T_h=20$, $T_r=4, \eta_1=\eta_2=0.2, \lambda_2=0.2$, $\beta=0.25, \epsilon=1$, and $M=5\times10^{-2}$. In Figs.  both of $\lambda_r=1$ and $\lambda_r=0.1$ are simulated to show the influence of the thermal conductivity ratio on the velocity and the temperature fields. It is shown in Fig. \ref{Fig5} that the velocity fields and the isothermal lines predicted by LBE are similar with that of analytical solutions \cite{Pendse}. For the quantitative comparison, in Figs. \ref{Fig6} and \ref{Fig7}, we also plot the velocity and temperature profiles across the center of the domain together with the analytical solutions along the $x$ and $y$ directions respectively. The results shown that numerical results agree with the analytical ones. Moreover, it is found that the lower value of $\lambda_r$ could strengthen the thermocapillary driven convection and lead to a more inhomogeneous temperature distribution along the interface. We also note that the difference of velocity components between numerical results and the analytical ones are increased as $\lambda_r$ reduced. The reason may be that the interface between two phases is a finite transition layer in LBE model, which implies that the fluid properties are variable across the interface.

\subsection{2D thermocapillary migration of deformable droplet}

In the limit of zero Marangoni number and small Reynolds number, Young et al \cite{Young} first analyzed the thermocapillary migration in an infinite domain with a constant temperature gradient $|\nabla T_\infty|$, and derived a theoretical expression for the migration velocity (also known
as YGB velocity) of a spherical droplet or bubble, which can be given as
\begin{equation}
U_{YGB}=\frac{2U}{(2+\lambda_r)(2+3\eta_r)}
\label{Eq42}
\end{equation}
where $U$ is the characteristic velocity defined by the balance of
the thermocapillary force and the viscous force on the droplet
or bubble as follows:
\begin{equation}
U=-\frac{\sigma_T|\nabla T_\infty|R}{\eta_2}
\label{Eq43}
\end{equation}
where $R$ is the radius of the bubble.

In our simulation, a planar 2D bubble of radius $R=20$ is initially placed inside a domain of size $8R\times15R$ with drop's center at the center of box and $2R$ above the bottom wall. Nonslip boundary conditions are imposed on the top and bottom walls, and periodic boundary condition is applied to the horizontal direction. A linear temperature field is imposed
in the vertical direction with $T=0$ on the bottom wall and $T=30$ on the top wall, resulting in $|\nabla T_\infty| = 0.1$. The model parameters are fixed as $\rho_1=\rho_2 =1$, $\epsilon=1, \beta=0.25$, viscosity $\eta_1=\eta_2=0.1$, thermal conductivity $\lambda_1=\lambda_2=0.1$, $T_0=0$, and $\sigma_0=0.1$. Then the theoretical bubble migration velocity of a spherical drop could be predicted by Eqs. (\ref{Eq42}) and (\ref{Eq43}) as $U_{YGB}=1.33\dot{3}\times 10^{-4}$. In the simulations, the migration velocity $u_r$ is calculated by
\begin{equation}
u_r(t)=\frac{\int_Vcu_ydV}{\int_VcdV}=\frac{\sum_{\bm x}cu_y}{\sum_{\bm x}c}, \text{where}~~c\geq0.5.
\end{equation}

\begin{figure}
\includegraphics[width=0.5\textwidth,height=0.38\textwidth]{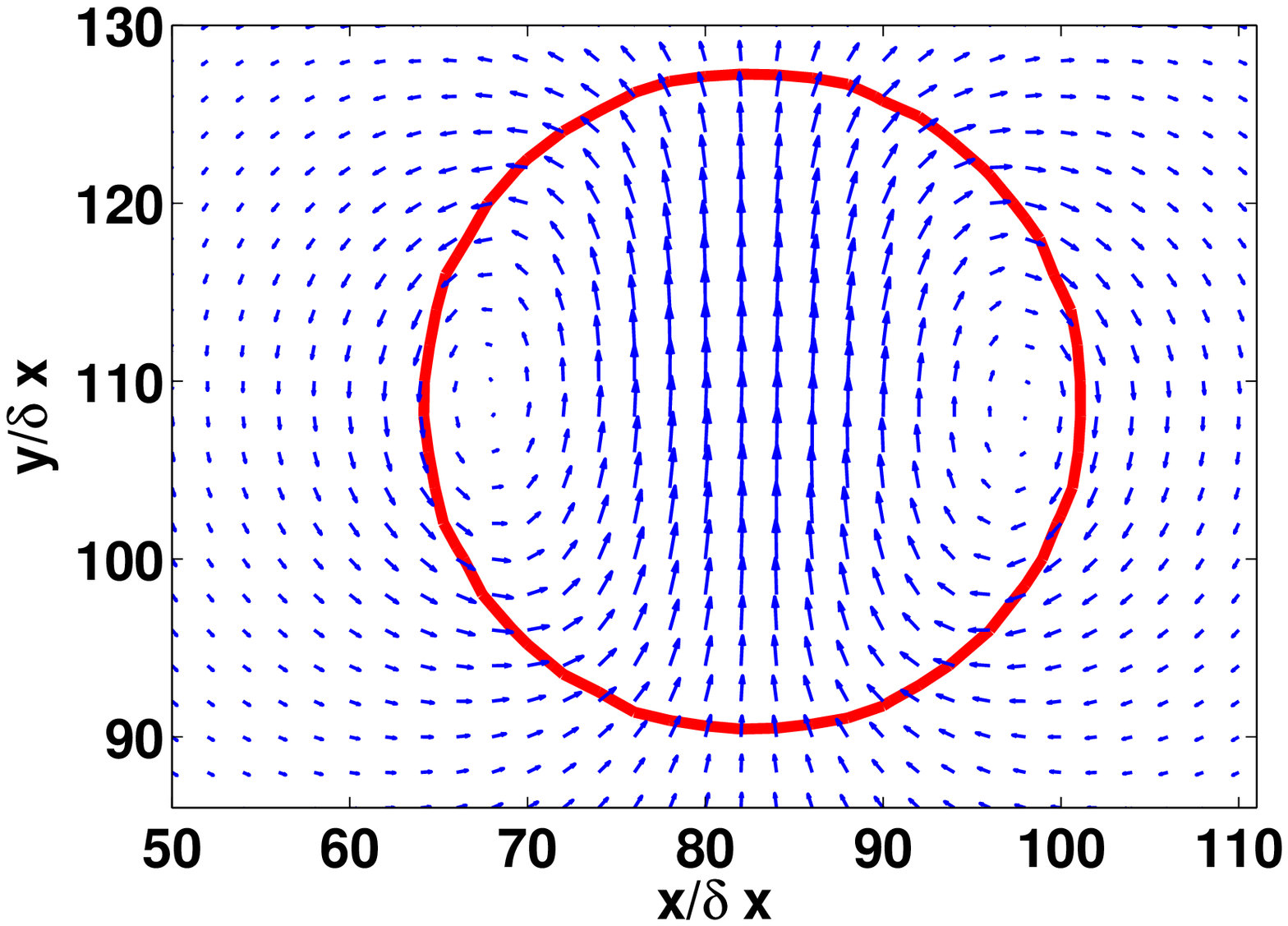}%
\includegraphics[width=0.5\textwidth,height=0.38\textwidth]{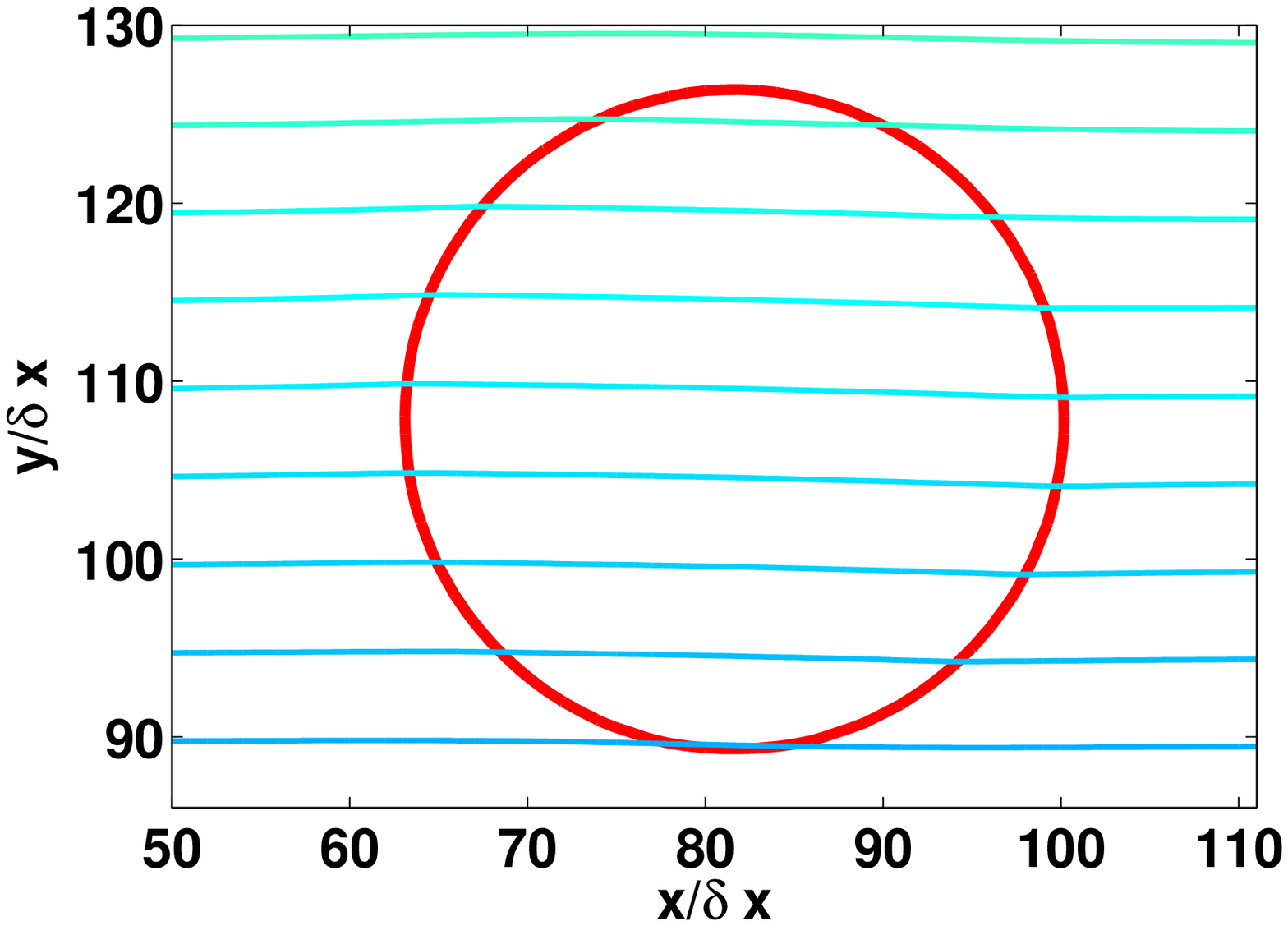}
\caption{Velocity vectors and temperature fields around the rising droplet}\label{Fig8}
\includegraphics[width=3in,height=2.5in]{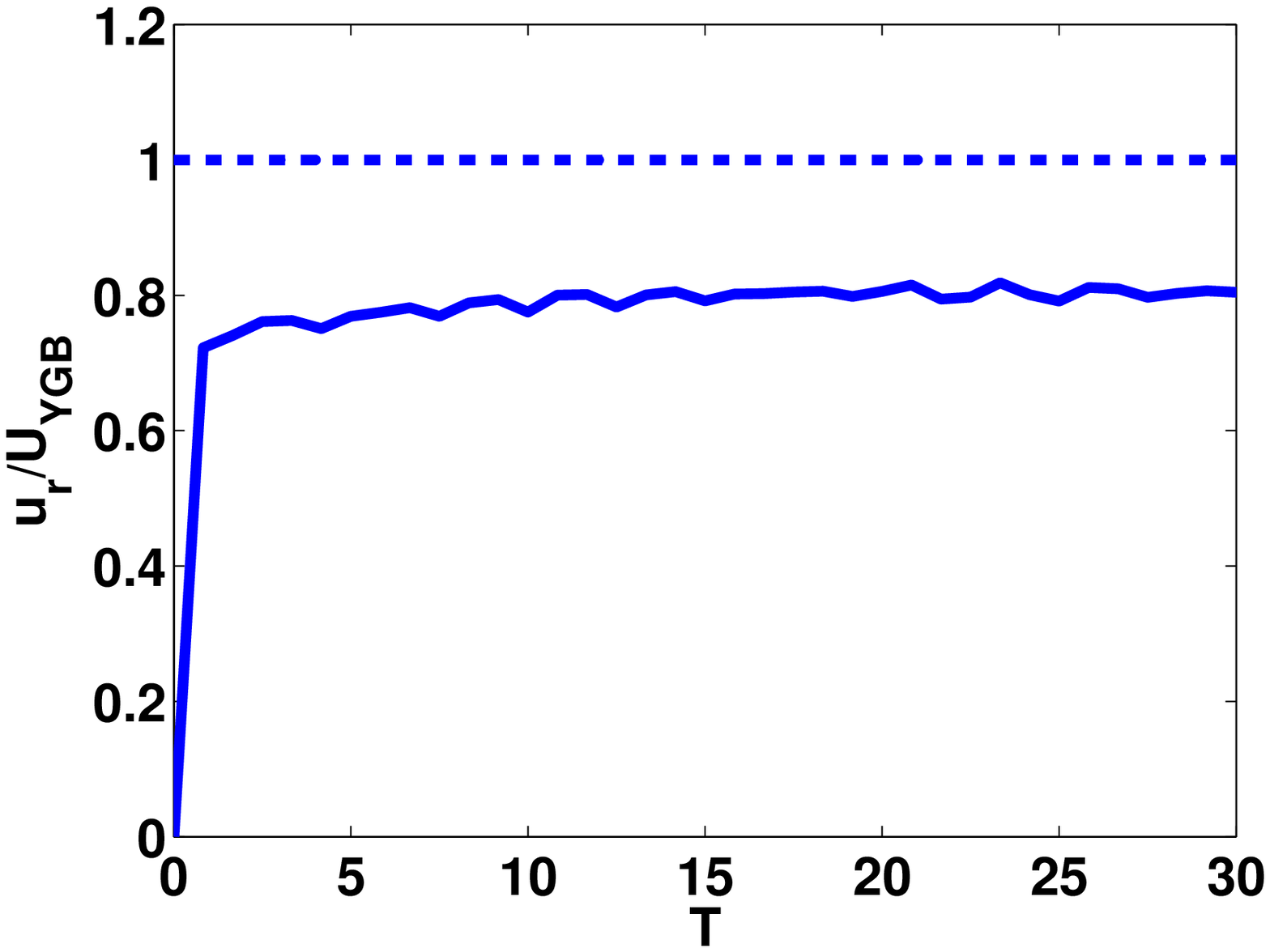}
\caption{Time evolution of normalized migration velocity of a droplet at Ma=Re=0.1. The dashed
line represents the theoretical prediction by Eq. (\ref{Eq42}), and the solid line is the numerical results}\label{Fig9}
\end{figure}

The velocity vectors and isothermal lines are shown in Fig. \ref{Fig8}, due to the variation of surface tension force visa the temperature gradient, the flow pattern within the droplet exhibits recirculation flow that make the droplet move from a cold region to a hot region, while the temperature is almost conduction throughout the droplet for such small Ma=0.1 and Re=0.1. To quantified our simulations, the temporal evolution of the migration velocity normalized by $U_{YGB}$ vs. dimensionless time $T$ normalized by $t_r=R/U$ is also shown in Fig. \ref{Fig9}. It is observed that the numerical result of the migration velocity $u_r$ seems to converge a value of $u_r/U_{YGB}\sim 0.8$, which implied the numerical prediction of planar droplet migration was roughly around $20\%$ off the theoretical one. The reason for this discrepancy is that the theoretical migration velocity is derived for an axisymmetric non-deformable sphere bubble/droplet in an infinite domain, while present simulations are conducted with a planar 2D deformable droplet in a finite domain. Similar observation of this phenomena was also obtained by the level-set method \cite{Herrmann} and phase field method \cite{Guozl}.

\section{Conclusion}
Continuous surface force (CSF) formulation was successfully used in traditional numerical methods such as level set, volume of fluid, and diffuse interface method. In this paper, we extended previous CSF LBE to the thermocapillary flow, the evolution of interface is governed by the Cahn-Hilliard equation which is solved by LBE, and a thermal LBE was derived from the kinetic theory for solving the scalar convection-diffusion equation. In literature, several LBE models have been designed to solve this equation, however, most of them can not derive the correct equation at macroscopic level.

Three benchmark tests including the thermal layered Poiseuille flow, two superimposed planar fluids at negligibly small Reynolds and Marangoni numbers for the thermocapillary driven convection, and a two dimensional deformable droplet migration by the temperature gradient are conducted to validate the present LBE. The results agree with analytical solutions and/or classic numerical predictions, this implied that present LBE could be applied to thermal multiphase systems.

\section*{Acknowledgments}

This work is supported by  the  Natural Science Foundation of Jiangsu Province (Grant No. BK20130750) and Research Fund for the Doctoral Program of Higher Education of China (20133219120047).

\appendix
\section*{Appendix A: Derivation of the energy equation}
\renewcommand{\theequation}{A\arabic{equation}}
\setcounter{equation}{0}

Through the Chapman-Enskog expansion technique, we can introduce the following expansions:
\begin{eqnarray}
&&\partial_t=\epsilon\partial_{t_1}+\epsilon^2\partial_{t_2}, \nabla=\epsilon\nabla, \bm{a}=\epsilon \bm{a}_1,\nonumber \\
&&h_i=h^{(0)}_i+\epsilon h^{(1)}_i+\epsilon^2 h^{(2)}_i.\nonumber
\end{eqnarray}

With above expansions and the equilibrium distribution function of Eq. (\ref{aEq9}),  Eq. (\ref{aEq8}) could be rewritten in consecutive orders of $\epsilon$ as
\begin{eqnarray}
\epsilon^0:h^{(0)}_i&=&h^{(eq)}_i,
\label{Aeq1}
\end{eqnarray}
\begin{eqnarray}
\epsilon^1:D_{1i} h^{(0)}_i&=&-\frac{h^{(1)}_i}{\tau_h}+F_i,
\label{Aeq2}
\end{eqnarray}
\begin{eqnarray}
\epsilon^2:~\partial_{t_2} h^{(0)}_i+D_{1i}h^{(1)}_i&=&-\frac{h^{(2)}_i}{\tau_h},
\label{Aeq3}
\end{eqnarray}
where $D_{1i}=\partial_{t_1}+\bm{\xi}_i\cdot\nabla_1$, and $F_i=\frac{\bm{\bar{F}}\cdot(\bm{\xi}_i-\bm u)}{\rho c^2_s}h^{(eq)}_i$.

With Eqs. (\ref{Aeq1})-(\ref{Aeq3}),  the macroscopic equations at $t_1$ scale can be derived
\begin{eqnarray}
\rho c_v(\partial_{t_1}T+u_\alpha\partial_{1\alpha}T)=0,
\label{Aeq4}
\end{eqnarray}

Similarly, the macroscopic equation at $t_2$ scale could be obtained as
\begin{eqnarray}
\rho c_v\partial_{t_2} T+\partial_{1\alpha}Q^{(1)}_{\alpha}&=&0,
\end{eqnarray}
where $Q^{(1)}_{\alpha}=\sum_i\xi_{i\alpha}h^{(1)}_i$.
To recover the corresponding macroscopic governing equation, the term of $P^{(1)}_{\alpha}$ should be estimated, which could be approximated by
\begin{eqnarray}
-\frac{1}{\tau_h}Q^{(1)}_{\alpha}&=&\partial_{t_1}(\sum_i \xi_{i\alpha}h^{(0)}_i)+\partial_\beta(\sum_i\xi_{i\alpha}\xi_{i\beta}h^{(0)}_i)-\sum_i \xi_{i\alpha}F_i\nonumber \\
&=&\partial_{t_1}(\rho c_vTu_\alpha)+\partial_\beta(\rho c_vTu_\alpha u_\beta+\rho c_vT c^2_s\delta_{\alpha\beta})-\sum_i \xi_{i\alpha}F_i\nonumber\\
&=& \rho c_vc^2_s\partial_\alpha T.
\label{Aeq6}
\end{eqnarray}
In the last step, we have used Eq. (\ref{Aeq4}) and the following relations
\begin{eqnarray}
\sum_i \xi_{i\alpha}F_i&=&c_vT\bar{F}_\alpha\nonumber\\
\rho(\partial_{t_1}u_\alpha+u_\beta\partial_\beta u_\alpha)&=&-\partial_\alpha\rho c^2_s+\bar{F}_\alpha\nonumber\\
\partial_{t_1}(\rho c_vTu_\alpha)+\partial_\beta(c_vTu_\alpha u_\beta)&=& \rho c_vT(\partial_{t_1}u_\alpha+u_\beta\partial_\beta u_\alpha) \nonumber
\end{eqnarray}
With aid of Eqs. (\ref{Aeq4})-(\ref{Aeq6}), the corresponding of the energy equation of Eq. (\ref{aEq14}) could be derived at macroscopic level.


\begin{thebibliography}{99}
\bibitem{Subra}R. S. Subramanian and R. Balasubramaniam, \emph{The motion of Bubbles and Drops in Reduced Gravity}, Cambridge University Press, 2001.
\bibitem{Darhuber} A. A. Darhuber and S. M. Troian, \emph{Annu. Rev. Fluid Mech.} \textbf{37}, 425 2005.
\bibitem{Young}N. O. Young, J. S. Goldstein, and M. J. Block, \emph{J. Fluid Mech.} \textbf{6}, 350 (1959).
\bibitem{Kawamura}H. Kawamura, K. Nishino, S. Matsumoto, and I. Ueno, \emph{Journal of Heat Transfer} \textbf{134}, 031005 (2012)

\bibitem{Sim}B-C. Sim, A. Zebib, \emph{Int. J. Heat Mass Transfer} \textbf{45}, 4983 (2002).
\bibitem{Hu}W. R. Hu, Z. M. Tang, K. Li, \emph{Applied Mech. Rev.}, \textbf{61}, 010803 (2008).
\bibitem{Du}R. Du, B. C. Shi, \emph{Computers \& Mathematics with Applications}, \textbf{55} 1433 (2008).
\bibitem{Zhenga}L. Zheng, Z. L. Guo, B. C. Shi and C. G. Zheng, \emph{AAMM}, \textbf{2} 677 (2010).
\bibitem{Hariri}H. Haj-Hariri, Q. Shi, and A. Borhan, \emph{Phys. Fluids} \textbf{9} 845, (1997).
\bibitem{Yin}Z. H Yin, L. Chang, W. R. Hu, Q. H. Li and H. Y. Wang, \emph{Phys. Fluids} \textbf{24}, 092101 (2012).

\bibitem{Liu}H. H Liu, Y. H. Zhang, and A. J. Valocchi, \emph{J. Comput. Phys.} \textbf{231}, 4433 (2012).
\bibitem{Liu1}H. H Liu, A. J. Valocchi and Y. H. Zhang, \emph{Phys. Rev. E} \textbf{87}, 013010 (2013).
\bibitem{He}X. He, S. Chen and R. Zhang, \emph{J. Comput. Phys.} \textbf{152}, 642 (1999).
\bibitem{Lee} T. Lee and C. L. Lin, \emph{J. Comput. Phys.} \textbf{ 206}, 16 (2005).
\bibitem{Zheng} L. Zheng, S. Zheng and Q. L. Zhai, unpublish.
\bibitem{Landau}L. D. Landau and E. M. Lifshitz, \emph{Fluid Mechanics}, Pergamon, New York, 1959.
\bibitem{BGK}P. L. Bhatnagar, E. P. Gross, and M. Krook, \emph{Phys. Rev.} \textbf{94}, 511 (1954).
\bibitem{He1}X. He, S. Chen, and G. D. Doolen, \emph{J. Comput. Phys.} \textbf{146}, 282 (1998).
\bibitem{Shi}Y. Shi, T. S. Zhao, and Z. L. Guo, \emph{Phys. Rev. E} \textbf{70}, 066706 (2004).
\bibitem{Pendse}B. Pendse, A. Esmaeeli, \emph{Int. J. Therm. Sci.} \textbf{49}, 1147 (2010).
\bibitem{Herrmann}M. Herrmann, J. M. Lopez, P. Brady, and M. Raessi, \emph{Proceedings of the Summer program}, 155 (2008).
\bibitem{Guozl}Z-L Guo and P. Lin, arXiv:1401.5793v3, (2014).

\end{thebibliography}
\end{document}